\newcommand{\fixed}[1] { #1 }

\documentclass[fleqn]{article}
\usepackage[headings]{espcrc2}

\usepackage{graphicx}
\usepackage[figuresright]{rotating}

\newcommand{\beq}[0] { \begin{eqnarray}}
\newcommand{\eeq}[0] { \end{eqnarray}}

\newcommand{\myfig}[2] {
\begin{figure}[htb]
 \vspace{9pt}
 \includegraphics[width=7.5cm]{#1.eps}
    \caption{#2} \label{#1}
    \end{figure} }

\newcommand{\myfigw}[2] {
\begin{figure*}[htb]
 \vspace{9pt}
\centerline{ \includegraphics[width=14cm]{#1.eps} }
    \caption{#2} \label{#1}
    \end{figure*} }

\title{
First-Principles Study on the Mobility of
Screw Dislocations in BCC Iron}

\author{M.Itakura
\address[JAEA-k]{
Center for Computational Science \& e-Systems, Japan Atomic Energy Agency.
5-1-5 Kashiwanoha, Kashiwa, Chiba 277-8587, Japan
}, 
H.Kaburaki
\address[JAEA-t]{Center for Computational Science \& e-Systems, Japan Atomic Energy Agency.
2-4 Shirakata-Shirane, Tokai-mura, Naka-gun, Ibaraki 319-1184, Japan}, 
and M.Yamaguchi\addressmark[JAEA-t]
} 

\begin{document}

\begin{abstract}
The fully two-dimensional Peierls barrier map of screw dislocations in 
body-centered cubic (bcc) iron 
has been calculated using the first principles method to identify the migration 
path of a dislocation core.
An efficient method to correct the effect
of the finite size cell used in the first-principles method
on the energy of a lattice defect
was devised to determine the accurate barrier profile.
We find that the migration path is close to a straight line that is confined in a $\{110\}$
plane and the Peierls barrier profile is single humped.
This result clarifies why  the existing 
empirical potentials of bcc iron fail to predict the correct mobility path.
A line tension model incorporating these first-principles calculation results
is used to predict the kink activation energy to be $0.73$ eV in agreement with experiment. 

\vspace{1pc}
\end{abstract}

\maketitle

\section{Introduction}
Recent nanopillar and nanowire experiments \cite{nano},
as well as in situ observation of dislocation motions \cite{caillard10},
clearly demonstrate
the highly peculiar natures of plasticity in body-centered cubic (bcc) metals.
Plasticity in bcc metals is mainly mediated by the thermal activation
of kink pairs in screw dislocation lines
owing to strong lattice friction,
and shows a strong dependence on the metal elements, the direction of the applied stress
and temperature \cite{spitzig70,vitek-bcc-review}.
In face-centered cubic metals,
the
lattice friction is extremely weak and a dislocation motion can be  described well
by the universal model of phonon drag.  
Therefore, a
reliable model of dislocation mobility in bcc metals
is indispensable for simulating plastic deformation in such metals.

One of the remarkable properties of a screw dislocation in bcc metals
is that its motion is
not confined in a plane
but varies with the orientation of the applied stress.
An improper model leads not only to qualitatively incorrect mobility
but also to a quantitatively wrong direction of migration.
Fig. \ref{fig01-coremig} shows  several possible migration paths
for a screw dislocation which moves from one stable ``easy-core''
position to another.
There are several unstable or metastable dislocation positions,
which are referred to as ``hard-core'' and  ``split-core'' positions.
Middle points between easy-core positions (M points)
are also shown in Fig. \ref{fig01-coremig}.

Anisotropic linear elasticity solutions (LES)
for each dislocation configuration
seen from the $[111]$ direction
are shown in  Fig. \ref{fig02-core-conf}
using the differential displacement map
\fixed{(see Appendix A)}.
The positive and negative numbers in the figure represents the helicity
of the dislocation core.
The solution for the M point is obtained by a simple interpolation
between the solutions of two adjacent easy-core configurations. This solution
can be  regarded as the sum of displacement fields induced by two ``half''
dislocation helicities located at the two easy-core positions. The solution
for the split-core configuration can then be obtained by linear extrapolation using
the hard-core configuration and the M point configuration. This solution can be
regarded as the sum of displacement fields induced by two positive dislocations
at the easy-core positions and one negative dislocation at the hard-core position.
Note that the LES for the 
easy-core and hard-core configurations
have $D_3$ point group symmetry in Fig. \ref{fig02-core-conf},
as is also true for the results of density functional theory (DFT) calculations for 
bcc iron and other bcc metals \cite{gfbc,bcc-dft,clouet09,gamma}.
We will later see that the displacement of the elasticity solutions 
for the split-core and M point
configurations is also close to that of the DFT calculations.

If either the hard-core or split-core configuration is metastable,
the migration of a dislocation
follows a bent line and the Peierls energy becomes double humped.
On the other hand, when neither of the intermediate configurations
is metastable, the path is close to  a straight line and
the Peierls energy becomes single humped.
The difference in this energy landscape strongly affects the 
kink nucleation energy of a screw dislocation in different slip directions,
and this
determines not only the average migration velocity but
also the average migration direction.
Indeed,
strong temperature and stress dependence on the average migration
direction of a screw dislocation is reported in
a molecular dynamics study of bcc iron screw dislocation \cite{md-slip-dir}.

However, since no quantum mechanical data of dislocation movement
are used for the fitting of the empirical potential of bcc iron employed in Ref. \cite{md-slip-dir},
the dislocation motion obtained in the molecular dynamics (MD)
simulations should critically be checked by the DFT-based simulations.
The Peierls energy in bcc iron without shear stress has been calculated by
Ventelon and Willaime
\cite{cea-dft-peierls}
using the DFT method with a localized basis.
In the present work
we employ the more accurate plane-wave-based DFT
to calculate a Peierls energy of an isolated screw dislocation core with and without
the applied shear stress, obtain a complete Peierls barrier map for the
two dimensional motion of a screw dislocation and identify its migration path.
By coupling the DFT result with the line tension model of a dislocation, we also    
estimate the average slip direction of screw dislocations for various
temperatures and applied stresses.

This paper is organized as follows.
We first describe the method used for the DFT calculations.
Further details of the DFT calculations to estimate the Peierls energy
are dealt with in Section 3.
In Section 4 we describe a method to estimate and correct the finite
size effect of DFT calculations using the Green's function method.
In Section 5 the results of the DFT calculations are summarized.
Section 6 presents the line tension model of a dislocation to estimate the 
kink pair nucleation energy
for various applied stresses.
Concluding remarks are given in Section 7.

\myfig{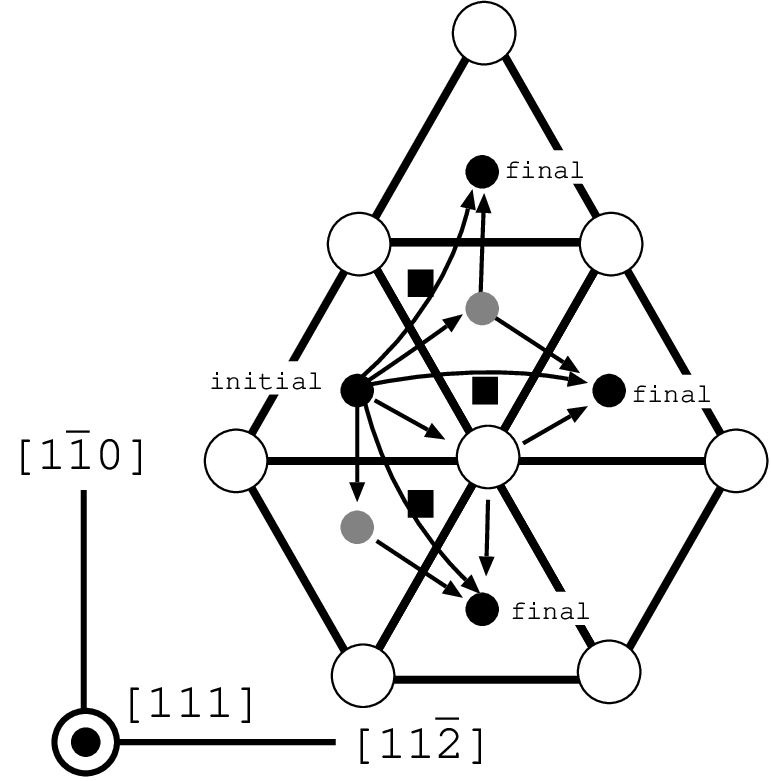}
{
Several possible migration paths of a screw dislocation in bcc metals.
Black,  gray, and white circles
represent the dislocation center positions
of easy-core, hard-core, and split-core configurations, respectively.
Black squares are middle points between easy-core positions (M points).
Metal atoms are located at the white circles. }

\myfig{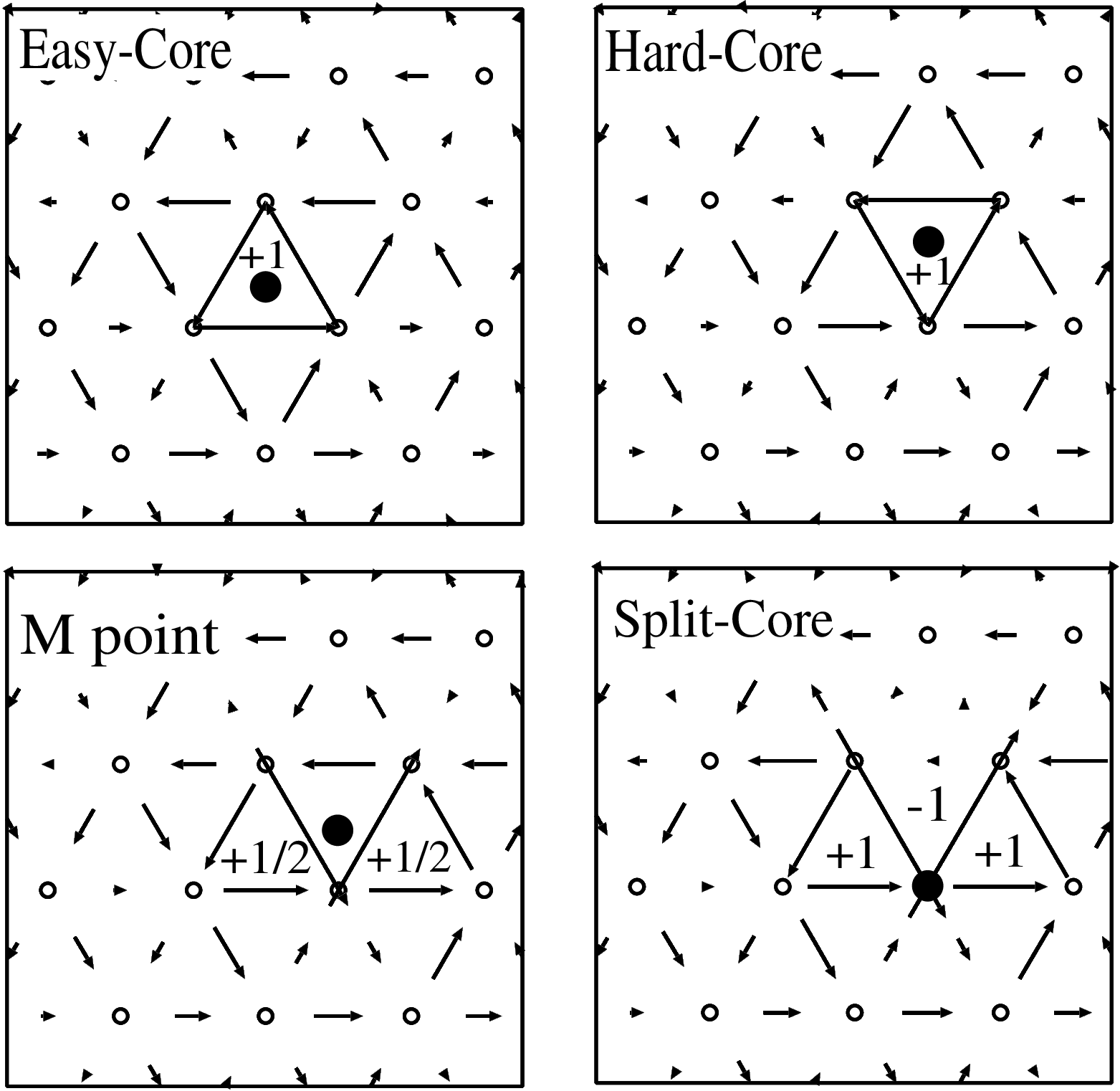}
{
Differential displacement maps of 
the linear elasticity solution for
easy-core, hard-core, M point, and split-core configurations.
Filled circles indicate the dislocation core position in each configuration;
the positive and negative
numbers indicate the dislocation helicity located at each triangle.
See the main text for details.
}

\section{The method of DFT calculations}

The electronic structure calculations and the structure
relaxations by force minimizations in the
DFT steps are performed using the Vienna ab initio simulation package,
 with the projector augmented wave method and ultrasoft pseudopotentials. 
The exchange correlation energy is calculated
by the generalized gradient approximation.
Spin-polarized calculations are employed in all cases.
The Methfessel-Paxton smearing method with 0.1-eV width is used.
Structural relaxation is terminated when the maximum force acting on the 
movable degrees of freedom
becomes less than 10 meV/$\AA$.
The error of energy that comes from this termination
is estimated to be $2$ meV by
comparing the energy with
the calculation for the smaller threshold of 3 meV/$\AA$.
The cutoff energy for the plane wave basis set is 400 eV, and
the convergence of Peierls barrier energy with respect to the increasing cutoff is confirmed.
The cutoff energy for the augmentation charges is 511.368 eV.

Periodic dislocation quadrupole configurations in
a parallelepiped cell shown in Fig. \ref{fig03-cell} defined by the following three edges,
\begin{eqnarray}
{\bf e_1}=& L_x {\bf v_{11\bar{2}}}, \nonumber \\
{\bf e_2}=&  \frac{L_x}{2} {\bf v_{11\bar{2}}} + L_y {\bf v_{1\bar{1}0}}+\frac{1}{2}{\bf v_{111}},\nonumber \\
{\bf e_3}=& {\bf v_{111}}
\end{eqnarray}
are used in all calculations,
where 
${\bf v_{11\bar{2}}}=a_0[11\bar{2}]/3$,
${\bf v_{1\bar{1}0}}=a_0[1\bar{1}0]$,
${\bf v_{111}}=a_0[111]/2$,
and
$a_0 = 2.833 \AA$ is a lattice parameter estimated by
a calculation of single BCC cell of
$[a_000]\times[0a_00]\times[00a_0]$ with 20x20x20 k-points.
DFT calculations are mainly carried out on the $L_x\times L_y = 15\times 9$ lattice , and 
the smaller and larger lattice sizes of
$L_x\times L_y = 9\times 5$, $9\times 7$, $15\times 7$, $21\times 11$ are also employed to
estimate the finite size effect.
Hereafter we introduce the Cartesian coordinates X, Y, and Z which are parallel to
the $[11\bar{2}]$,  $[1\bar{1}0]$, and $[111]$ directions, respectively.
Two dislocations with opposite helicity are placed in the cell to
form a periodic quadrupolar array. 
\myfigw{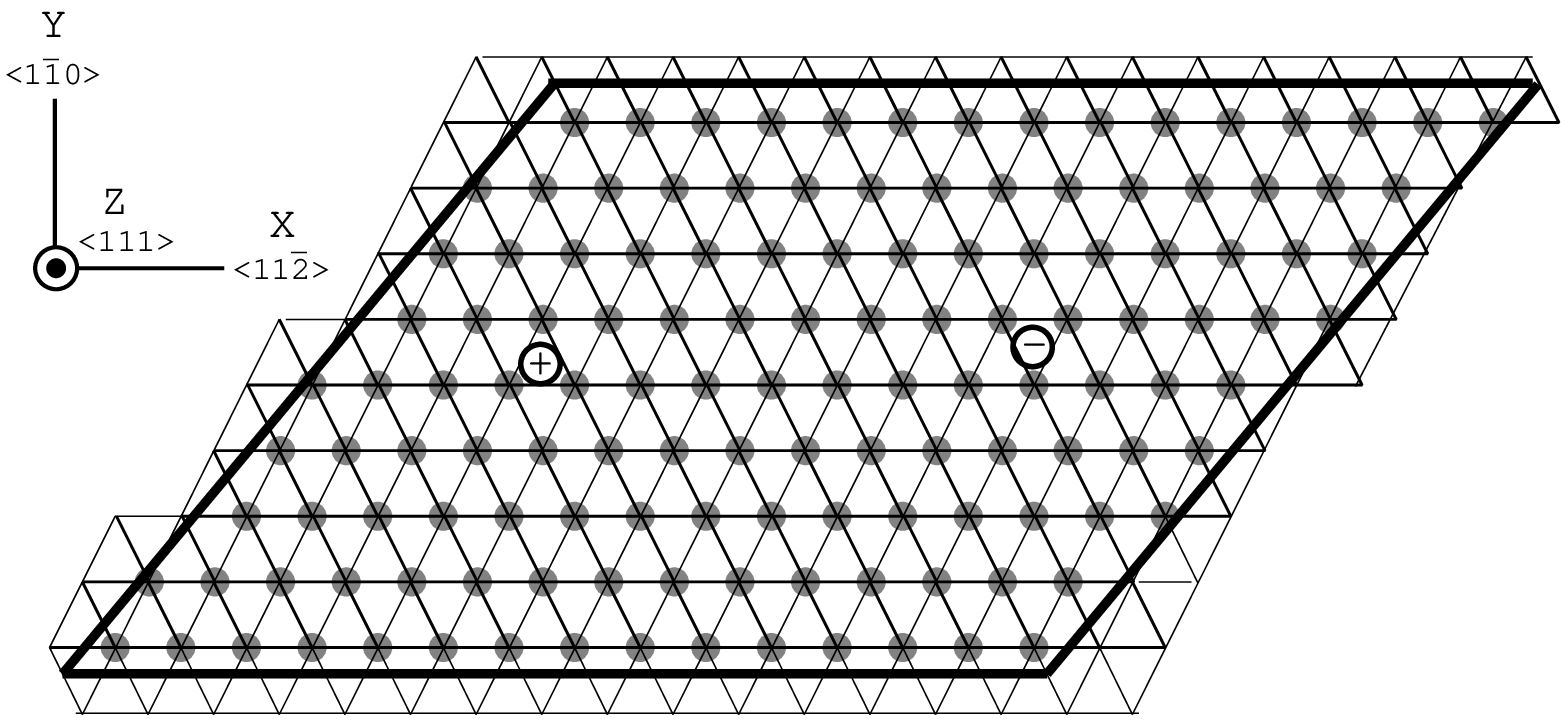}
{
Periodic quadrupolar array of dislocation cores
in a parallelepiped cell used in the DFT calculations, seen from the $[111]$ direction. 
Gray circles are iron atoms and the two white circles are the two dislocations
with opposite helicities.
}

The reciprocal lattice vectors of the parallelepiped described above are
$(k_x, -k_y/2, 0)$, $(0, k_y, 0)$, and $(0, -k_y/2, k_z)$ where
$k_x=\sqrt{6}\pi/(2a_0L_x) $, $k_y=\sqrt{2}\pi/(2a_0L_y)$, and $k_z=2\sqrt{3}\pi/(3a_0)$.
Since the Monkhorst Pack k-point
mesh for this inclined Brillouin zone breaks the symmetry of
the independent inversions in the $X$, $Y$, and $Z$ directions,
we use a Cartesian k-mesh of size $k_x/n_x \times k_y/2n_y \times k_z/16$,
which preserves
the symmetry. The mesh number $n_x$ and $n_y$ are adjusted so that the mesh size becomes close
to $0.1 \AA^{-1}$.
Convergence of the Peierls barrier energy 
with respect to the increasing mesh number  is confirmed.

\section{Calculation of the Peierls Barrier}
We denote the displacement of each atom relative to the perfect crystal position by
$\vec{u}_i=(u_i^x, u_i^y, u_i^z)$ where $i$ is an index of an atom.
We also define a differentiated displacement
as the normalized difference of the 
Z-component displacement for each pair of neighbor atoms
$u_{ij}=(u_i^z-u_j^z)/b$, where $b$ denotes the length of the
Burgers vector.
\fixed{For both $\vec{u}$ and $u_{ij}$, the minimum image convention
in the $z$ axis is applied, and $|u_{ij}| \leq 1/2$ is always satisfied.}
We define the two dimensional Peierls energy as
the minimum energy under the constraint of dislocation location
being at a certain position in the two dimensional plane.
By watching the differentiated displacement $u_{ij}$, one can determine
in which triangle a dislocation is located: a directed sum of three $u_{ij}$ around
a triangle is equal to the dislocation helicity inside it.

It is easy to show that a set of possible values of three $u_{ij}$s around 
a triangle which has non-zero dislocation helicity becomes an equilateral
triangle in the three-dimensional space of $u_{ij}$. Thus it would be natural to
map this triangular region of $u_{ij}$ to the triangle in the real space
which surrounds the dislocation to define the intra-triangular
dislocation position.
Its explicit definition is given as:
\begin{equation}
\vec{P}=-2( \vec{r_1} u_{32} +\vec{r_2} u_{13} + \vec{r_3} u_{21}) 
\label{DislPos},
\end{equation}
where $\vec{r_i}$ is the two-dimensional position of vertex $i$ 
relative to the center of the triangle, as shown in 
Fig. \ref{fig04-trimap}. The position is relative to  the center of the triangle.
The Cartesian coordinate of the position is given as:

\begin{eqnarray}
P_x =& \frac{2 \sqrt{2}}{3} (2 u_1^z-u_2^z-u_3^z), \nonumber
P_y =& \frac{2 \sqrt{6}}{3} (u_2^z - u_3^z). \label{pxpy}
\end{eqnarray}

For a given dislocation position $(P_x, P_y)$,
we define the Peierls energy $E_P(P_x,P_y)$  as follows:
\begin{eqnarray}
E_P(P_x,P_y)&=&E_d(x^d)= \min_{x^o} E(x^d, x^o) \nonumber \\
 &=& E(x^d, \bar{x}^o(x^d))
 \end{eqnarray}
where $x^d$ denotes a set of degree of freedom that is related to the dislocation
position and is determined by $(P_x, P_y)$.
The set of all the other degrees of freedom is denoted by $x^o$.
In the present case, $x^d$ are the $Z$ displacements of the three atoms around a dislocation. 
The energy $E_d(x^d)$ is a minimum of the total energy $E(x^d, x^o)$ for a given
value of $x^d$, and $\bar{x}^o(x^d)$ gives the values of $x^o$ 
which minimizes the energy.

In the DFT calculations or MD simulations,  $E_d(x^d)$ can be easily obtained by
fixing the values of $x^d$ and relaxing all the other degrees of freedom.
Moreover, one can calculate the derivative of $E_d(x^d)$ with respect to 
one of the fixed degrees of freedom $x^d_i$
from the forces acting on $x^d_i$ which are also easily obtained, as follows:
\begin{eqnarray}
\frac{\partial E_d(x^d)} {\partial x^d_i} 
& =& \frac{\partial E(x^d, \bar{x}^o(x^d))}{\partial x^d_i} \nonumber\\
& =&  \frac{\partial E(x^d, x^o)}{\partial x^d_i}
+  \frac{\partial E(x^d, x^o)}{\partial x^o} \frac{\partial \bar{x}^o(x^d)}{\partial x^d_i}
\nonumber \\
& =& -F_i,
\end{eqnarray}
where $F_i$ is just a force acting on $x^d_i$. The term $\partial E(x^d, x^o)/\partial x^o$ is zero
 at $\bar{x}^o$ because it gives minimum of $E(x^d, x^o)$ with respect to $x^o$.
The force acting on the dislocation is calculated as follows:
\begin{eqnarray}
-\frac{\partial  E_P}{\partial P_x}=&\frac{\sqrt{2}}{8} (2 F_1^z -F_2^z-F_3^z),\nonumber \\
-\frac{\partial  E_P}{\partial P_y}=&\frac{\sqrt{6}}{8} (F_2^z-F_3^z).
\end{eqnarray}

It should be noted that the core position does not change when
the three atoms move in the $Z$ direction by the same amount $d$,
and from the translational symmetry, $E_d$ also remains unchanged.
When there are two or more dislocations in the system separated by a
long enough distance, energy difference
induced by  independent translations of $x_d$ for each dislocation mainly comes
from the long-range elastic interaction between two dislocations, and is
minimized by using relative relation of $x_d$, which is given by a linear
elasticity solution.

\myfig{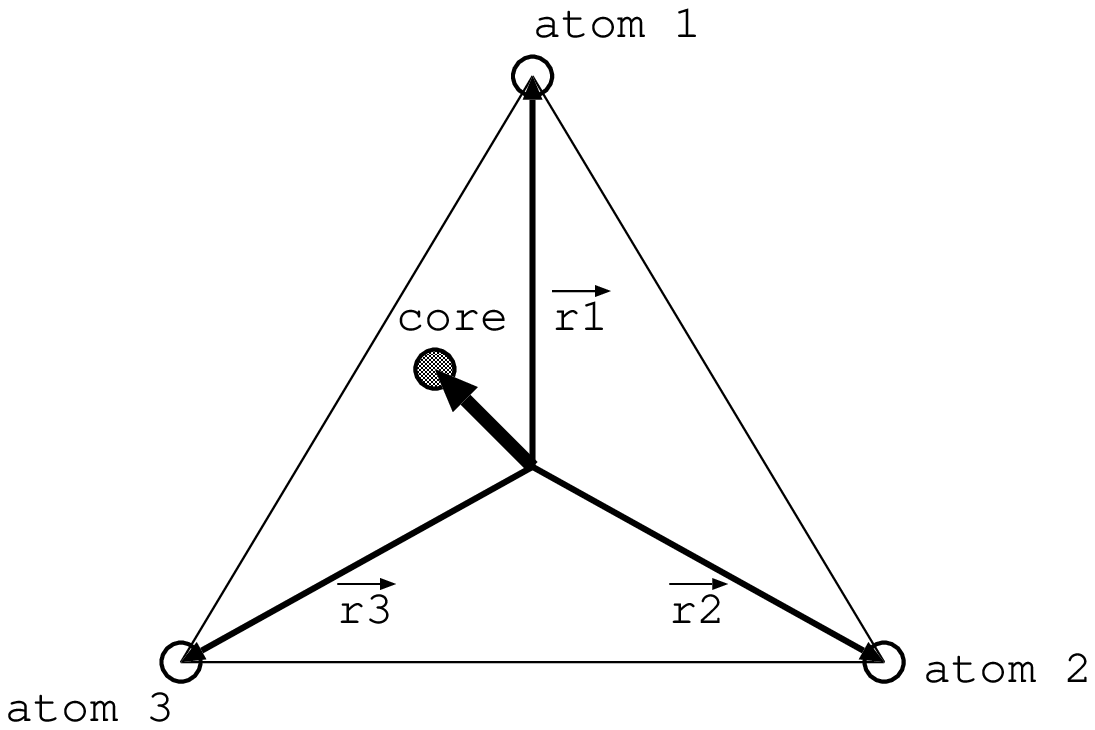}
{
Dislocation core position given by
$-2(\vec{r_1} u_{32} + \vec{r_2} u_{13} +  \vec{r_3} u_{21})$,
where $\vec{r_i}$ is the position of vertex $i$ relative to the center of the triangle.
See the main text for details.
}

In general, Peierls energy $E_P$ as a function of dislocation position $P$
is discontinuous at the boundary between two triangles,
since the choice of $x_d$ differs in each triangle and 
different values of $\bar{x}^o$ are obtained for different minimization condition.
To deal with this discontinuity, we extend the mapping domain of
Eq. (\ref{DislPos}) in a hard-core triangle 
to include three adjacent easy-core positions, as shown in Fig. \ref{fig05-hexmap}.
The values of $u_{ij}$ which correspond to these
three extended vertexes becomes
$u_{ij} =\pm 1/3$ on the boundary edge between the current hard-core
and adjacent easy-core triangle and $u_{ij} =\mp 1/6$ 
on the other two edges.
(Note that when the dislocation position crosses over 
a boundary between two triangles, $u_{ij}$ on the boundary
edge discontinuously changes from $1/2$ to $-1/2$ if the
minimum image convention is used. Instead, we allow $u_{ij}$
to become larger than $1/2$ outside the original triangle
so that the position given by Eq. (\ref{DislPos}) becomes continuous.)
These  coincide with the values of $u_{ij}$ when the dislocation is at
the center of the adjacent easy-core triangle
if the easy-core state is $D_3$ symmetric as is the case of bcc iron.

Therefore, by controlling the $Z$ displacement of the three atoms around a
hard-core triangle and relaxing all the other degrees of freedom,
we can obtain a hard-core configuration, three adjacent easy-core configurations, and
any intermediate configurations between them.
A continuous energy landscape of an area
which covers all the relevant  configurations
for a dislocation core migration is also obtained.
Later we will see that the split-core configuration has a very high energy
according to the first-principles calculation,
in contrast to the MD empirical potential calculation, and the migration path does not approach
the split-core vertexes in the hexagon but always remains inside the hexagon.

From the symmetry, the energy landscape inside a hexagon can be
determined by investigating the energy inside a triangle
the three vertexes of which are easy-core, hard-core, and split-core positions.
Peierls energies for the intra-triangular positions denoted by P0 to P18
shown in Fig. \ref{fig06-dftpos} are calculated using the DFT,
and the two dislocations are set to the same intra-triangular position.

\myfig{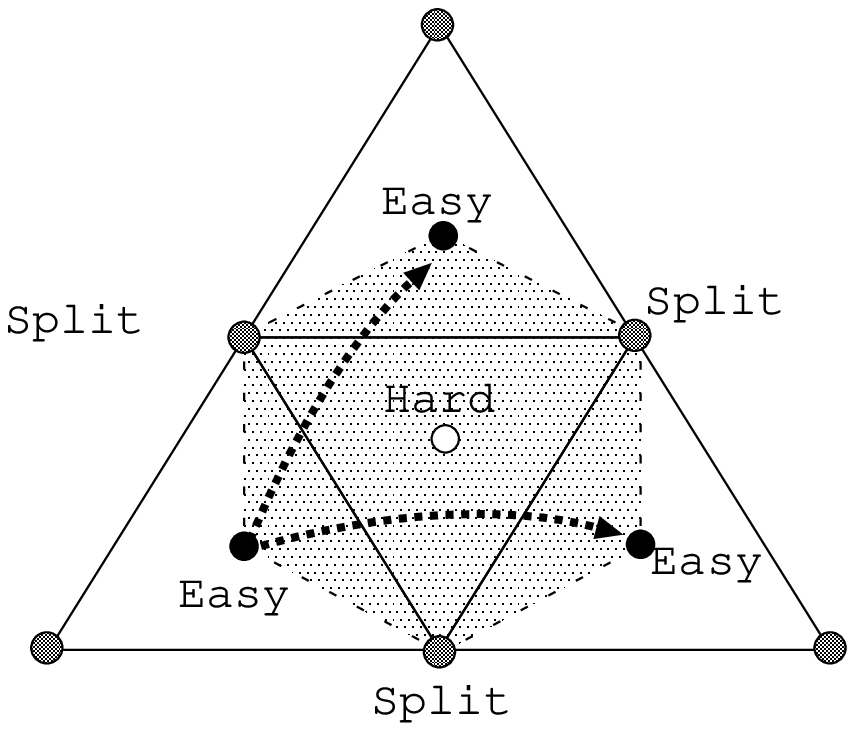}
{
The domain of mapping between the core position and the Z-displacement of three atoms around
a hard-core position is extended to a regular hexagon which includes three adjacent
easy-core positions.
}
\myfig{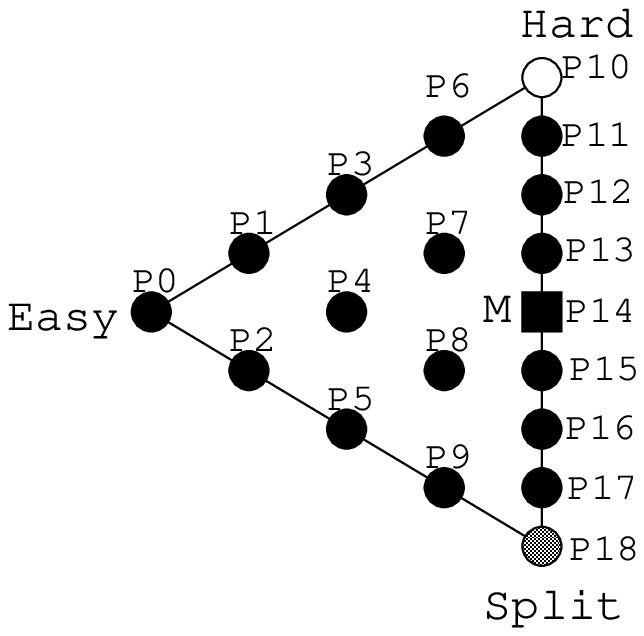}
{
The intra-triangular positions P0 to P18 used in the DFT calculations.
}

\section{Finite Size Effects}
Clouet et al. \cite{corefield} showed that
the difference in the dislocation displacement field
between LES and the DFT calculation 
can be described well by the  line forces near the core, which come
from the heavily displaced atoms at the core region.
While this ``core force'' is very localized at the core region,
the displacement owing to this force, referred to as the core field,
is inversely proportional to  the distance
from the core, and the interference between
the core fields induced by the multiple dislocations and their
mirror images
results in finite size corrections of various quantities.
In the present work we directly observe the core force
using the DFT calculation
for configurations given by LES, and
calculate the core field by the lattice Green's function method \cite{gfbc}
to estimate the 
finite size effect.

Figure \ref{fig07-fc} shows forces acting on each atom, 
$F^c_i$,
calculated by DFT for the case of LES of
the easy-core, hard-core,  split-core and  M position configurations
in the $21\times 11$ system. The core force for the smaller case of $15\times 9$ is
very close to the $21 \times 11$ case, and the maximum difference of forces between
them is $0.02$ eV/$\AA$.
In the easy-core configuration, the distance between the innermost three atoms and
the second innermost atoms is about 96\% of that of a perfect crystal, and
the second innermost atoms are pushed outward.
In the hard-core configuration, the distance between the innermost three atoms
is about 94\% of  that of a perfect crystal, and the innermost atoms
are pushed outward more strongly than in the easy-core case.

The DFT energy of the LES, 
$E_{LES}^{DFT}$, also shows a very weak size dependence.
Let us consider the relative value of $E_{LES}^{DFT}$ 
with respect to the easy-core configuration divided by the number of dislocation,
$\Delta E_{LES}^{DFT}$.
Hereafter the symbol $\Delta$ is also used for other quantities
with the same meaning.
This difference of energy mainly comes from the energy difference of 
core structure $\Delta E_c$ and the difference of the interaction energy between 
dislocations $\Delta E_{INT}$. We assume that 
$\Delta E_{LES}^{DFT}= \Delta E_c+\Delta E_{INT}$ and
$ E_{INT}$ is given by
\beq
\frac{\mu b^3}{2\pi}\left( -\log(r_{12})
+ \sum_{i=1,2} \sum_j q_i q_j \log(r_{ij})\right),
\eeq
where $\mu=57$ GPa is the shear modulus of the strain in $\{1\bar{1}0\}<111>$  calculated by DFT,
$i$ and $j$ denote the index of dislocation inside the cell and all the outside mirror images,
respectively,
$q_i=\pm 1$ is 
the helicity of the dislocation i and $r_{ij}$ is the distance between dislocation i and j.
Fig. \ref{fig08-ene-raw} shows dependence of $\Delta E_c$ of the hard-core, M point and split-core
configurations on the system size $L_x \times L_y$. 
Solid and dashed lines correspond to the data of the systems with
the aspect ratio $\sqrt{3} L_y/L_x$ being greater and less than 1, respectively.
These two cases show the opposite size dependence except for the 
split-core case in the $9\times 7$ system size.
From this result,
the finite size effect for $\Delta E_c$ 
at the size $15\times 9$ is estimated  
to be about $4$ meV.

\myfigw{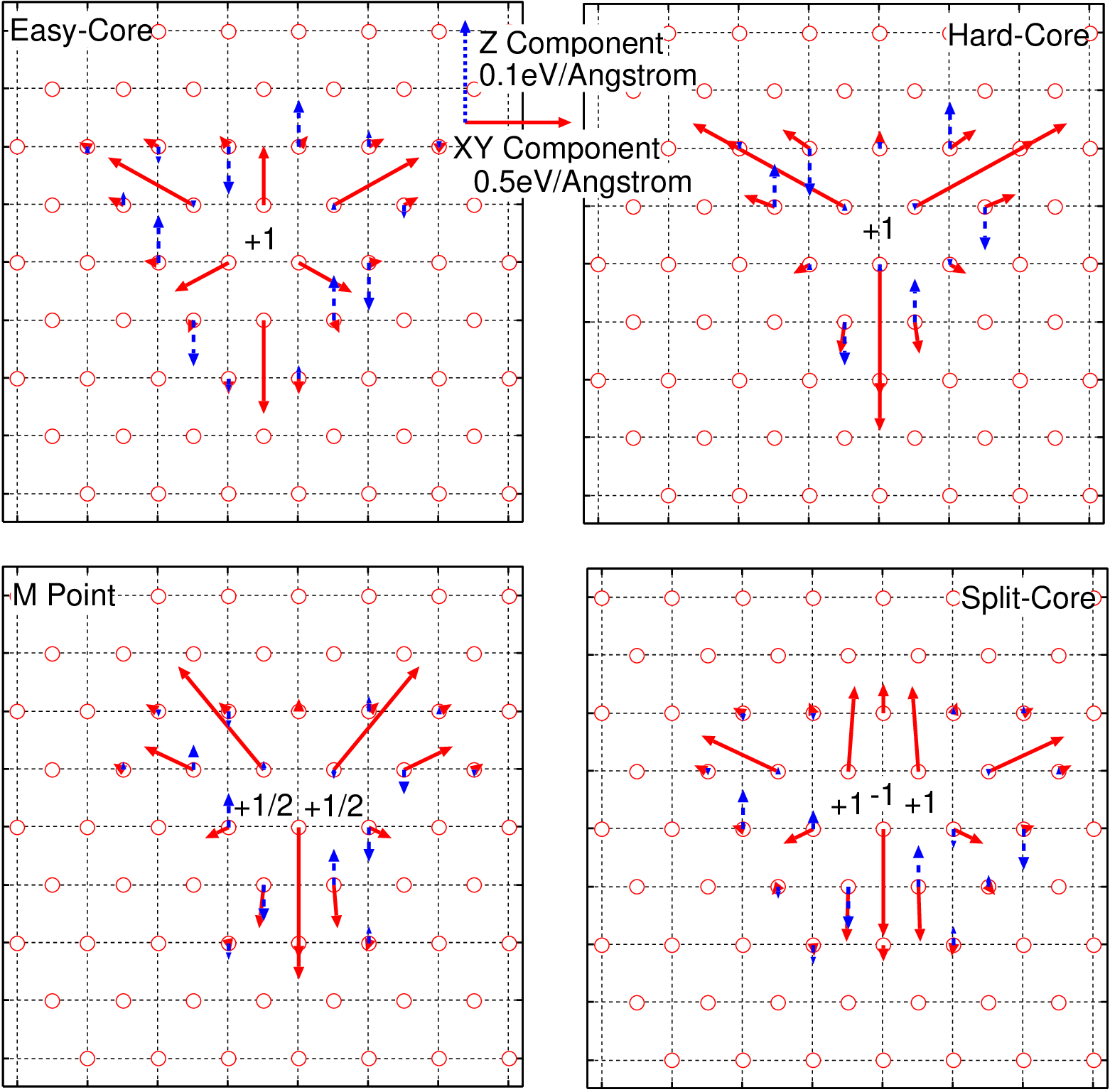}
{
 Core forces of the easy-core, hard-core, split-core and M position configurations.
Z-components of the force
are shown by vertical arrows and exaggerated five times compared to
the XY components.
Forces less than $0.05$ eV/$\AA$ are omitted for clarity.
}

\myfig{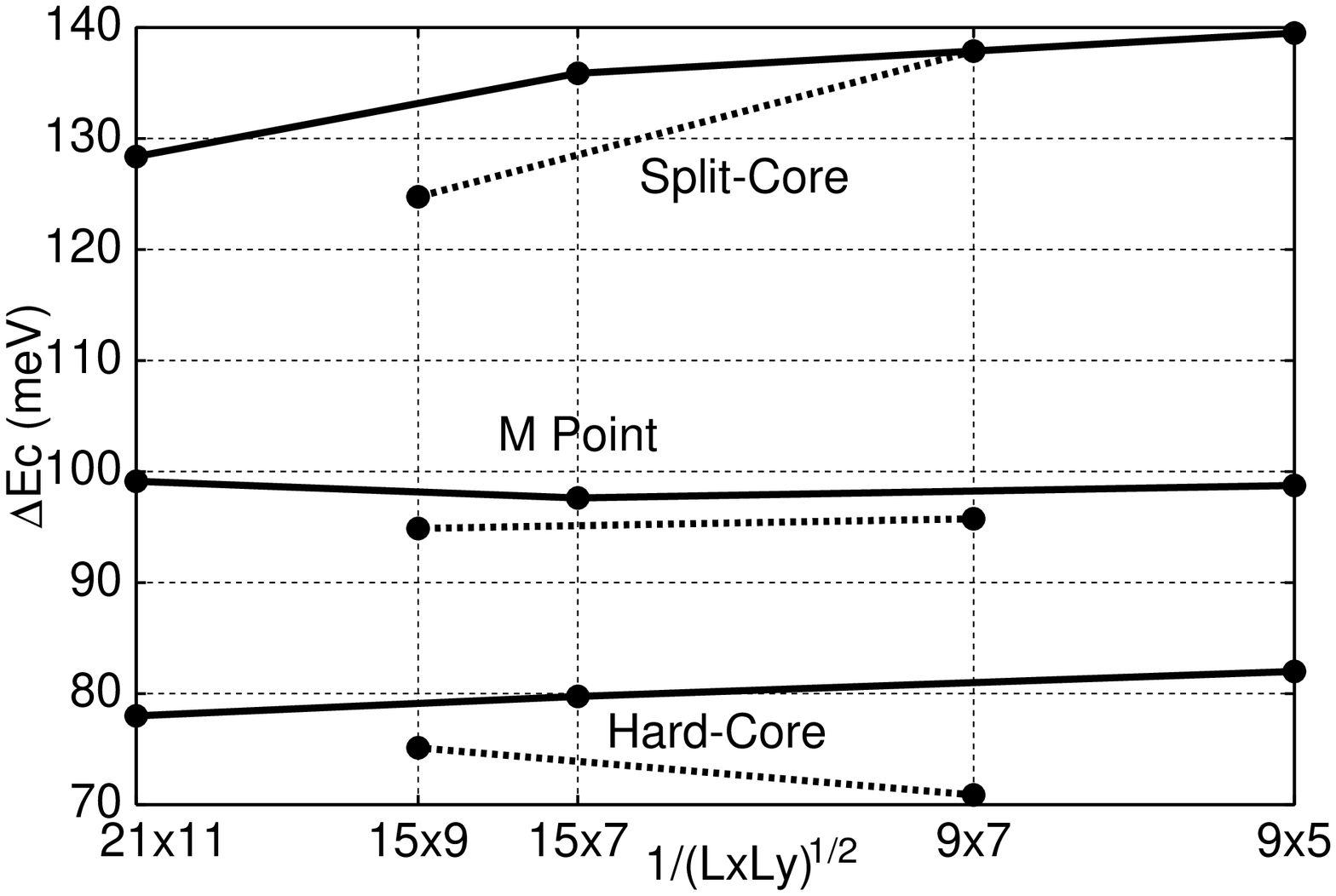}
{
The size dependence of the core energy $\Delta E_c$
of the hard-core, M point and the split-core configurations.
Solid and dashed lines correspond to the data of the systems with
the aspect ratio $\sqrt{3}L_y/L_x$ being greater or less than 1, respectively.
}

The DFT energy of the relaxed configuration is denoted by $E^{DFT}$,
and the energy difference $E^{DFT} -E^{DFT}_{LES}$ is denoted by $E_{RLX}^{DFT}$.
This relaxation energy
can be approximated by $-\frac{1}{2} \sum_i F^c_i u^c_i$,
where  $u^c_i$ denotes the change of displacement by the relaxation.
This change of displacement can be approximately calculated by
the linear elasticity theory, and in the present work 
we use the lattice Green's function method.
We assume that a small perturbation $\delta$ to a degree of freedom $j$
induces a force $H_{ij} \delta$ on the degree of freedom $i$,
and that the Hessian matrix $H_{ij}$,
which is the inverse matrix of the lattice Green's function,
is  approximated well by that of a perfect
crystal.
Then the core field $u^c_j$ can be obtained by solving the following
linear equations;
\beq
\sum_j H_{ij} u^c_j = -F^c_i \label{greenfunc}.
\eeq

To calculate $H_{ij}$,
a perfect crystal system 
is prepared using
the same system size and boundary condition as
those for
the dislocation calculation,
and one atom is displaced in each of the
$X$, $Y$, and $Z$ directions.
The forces induced on each atom is then observed using the DFT calculation
with the same parameters as the dislocation calculations.
We find that 
the induced forces 
are proportional to the amount of displacement up to $0.08 \AA$,
and that the force
is only significant at the displaced atom and its nearest neighbors, while
forces on the other atoms are negligible.
If we denote by $H_{ij}^{ab}$ the matrix element for 
the displacement of the atom $i$ in the direction $a$ 
and the displacement of the atom $j$ in the direction $b$,
the observed values are $H_{ii}^{XX}=H_{ii}^{YY}=-10.6$ eV/$\AA^2$,
$H_{ii}^{ZZ}=-6.03$ eV/$\AA^2$, and $H_{ii}^{ab}=0$ for all $a\neq b$ cases.
For the neighbor atom $j$ located at the $+X$ direction of the atom $i$
and displaced by $-b/3$ in the $Z$ direction compared to the atom $i$,
the matrix is $H_{ij}^{XX}=3.32$ eV/$\AA^2$,
$H_{ij}^{YY}=0.201$eV/$\AA^2$, 
$H_{ij}^{ZZ}=1.00$eV/$\AA^2$,
$H_{ij}^{XZ}=H_{ij}^{ZX}=0.331$eV/$\AA^2$,
and $H_{ij}^{XY}=H_{ij}^{YX}=H_{ij}^{YZ}=H_{ij}^{ZY}=0$.
Matrix elements for other neighbor atoms can be calculated from these values
using the rotational symmetry.
 The elastic constants expressed in the cubic axis
 derived from this matrix are
$C_{11}= 237$ GPa, $C_{12}= 104$ GPa, and $C_{44}=116$ GPa,
 which are in reasonable agreement with the experimentally observed
 values $C_{11}=243$, $C_{12}= 145$, and $C_{44}= 116$ GPa.
 The calculated anisotropy ratio $2C_{44}/(C_{11}-C_{12})=1.74$ 
 indicates that the elastic anisotropy is included in the matrix $H_{ij}$,
 although the ratio is smaller than the experimental value $2.36$.

The relaxation energy can then be estimated from the lattice Green's function as 
follows:
\beq
E_{RLX}^{GF} = -\frac{1}{2} H^{-1}_{ij} F^c_i F^c_j.
\eeq
If $E_{RLX}^{GF}$  agrees well with the actual value $E_{RLX}^{DFT}$,
one can estimate the relaxation energy of an isolated dislocation,
$E_{RLX}^{GF\infty}$,
using the core force of only one dislocation and using a sufficiently large lattice
in the calculation of $u^c_i$, and also estimate the finite size effect 
$E_{FSE}^{GF}= E_{RLX}^{GF\infty} - E_{RLX}^{GF}$.
Finally, the Peierls energy $\Delta E_P$
for an isolated dislocation is estimated as
$\Delta E_P = \Delta E^{DFT} -\Delta E_{INT} + \Delta E_{FSE}^{GF}$.

\section{Results}
Table \ref{tab:ene} summarizes
energies of the various core positions shown in 
Fig. \ref{fig06-dftpos}, calculated using the $15\times 9$ system.
For the calculation of $E^{GF}_{RLX}$, core forces smaller than $0.05$ eV/$\AA$
are omitted, and a periodic rectangle of $150\times 90$ atoms system is used 
to calculate $E^{GF\infty}_{RLX}$.
The agreement between $E^{DFT}_{RLX}$ and $E^{GF}_{RLX}$ is excellent, 
and we can  use the Green's function method to
estimate the finite size effect on the Peierls energy. 
Our estimate of the finite size effect is $2$ meV for positions near the 
M point and the hard-core position, and $5$ meV near the split-core position.
\fixed{
Together with the $4$ meV uncertainty of the finite size effect on $\Delta E_c$,
the numerical error for the Peierls potential is estimated to be
    $6$ meV for P9, P16, P17 and P18, and
    $5$ meV for the other cases.
}
Fig. \ref{fig09-ucall} shows a comparison of the core fields
between the DFT calculations and Green's function method for the case
of easy-core, hard-core, M point and split-core configurations.
The agreement is excellent, except near the core region.
Considering that the finite size effect mainly comes from
the  long-range part of the core field, we expect that 
the error of the core field near the core region
calculated by the lattice Green's function method
does not affect the calculation of finite size effect $E_{FSE}^{GF}$.

\myfigw{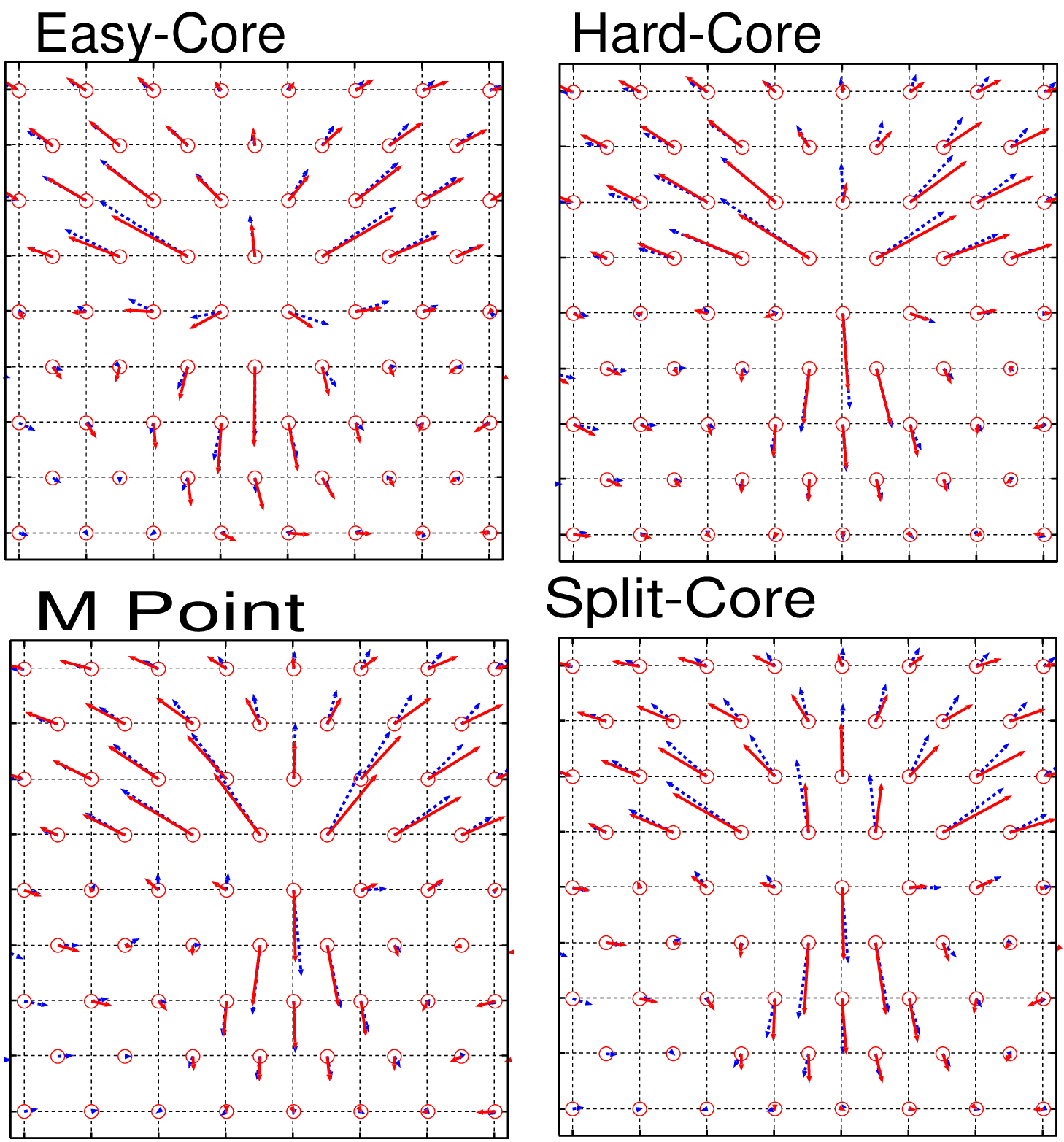}
{
A comparison of the core fields calculated by
the  DFT (solid red arrows) and the
lattice Green's function method (dashed blue arrows),
respectively. Only the XY components are shown for clarity, and they are exaggerated
by a factor of 40.
}

\begin{table*}[htb]
\renewcommand{\arraystretch}{1.2} 

\caption{
\label{tab:ene}
DFT energies and calculated correction terms for
various core positions
for the system size $15 \times 9$. 
All values are in the unit of meV.
\fixed{ The numerical error for the
Peierls energy $\Delta E_P$ is estimated to be
$6$ meV for P9, P16, P17 and P18,
and $5$ meV for the other cases.
See the main text for  details.
}
}

\begin{tabular}{ccccccccc}
\hline
P & $\Delta E_{LES}^{DFT}$ & $\Delta E^{DFT}$ & $E^{DFT}_{RLX}$
& $E^{GF}_{RLX}$ & $E_{RLX}^{GF\infty}$ & $\Delta E_{FSE}^{GF}$
& $\Delta E_{INT}$ & $\Delta E_P$ \\
\hline
P0 & 0 & 0 & -215 & -207 & -227 & 0.0 & 0.0 & 0.0 \\
P1 & 16 & 3 & -240 & -227 & -247 & -0.1 & 0.1 & 3.2 \\
P2 & 19 & 11 & -230 & -218 & -238 & 0.1 & -0.2 & 11.5 \\
P3 & 52 & 19 & -280 & -270 & -289 & 0.2 & 0.2 & 19.2 \\
P4 & 40 & 19 & -258 & -248 & -270 & -0.9 & 0.0 & 17.6 \\
P5 & 56 & 41 & -244 & -240 & -265 & -2.0 & -0.5 & 39.9 \\
P6 & 83 & 31 & -319 & -311 & -329 & 0.7 & 0.1 & 31.1 \\
P7 & 69 & 30 & -292 & -284 & -305 & -0.4 & 0.1 & 29.9 \\
P8 & 65 & 41 & -264 & -260 & -285 & -2.5 & -0.2 & 38.7 \\
P9 & 93 & 79 & -242 & -245 & -274 & -4.5 & -0.8 & 75.2 \\
P10 & 95 & 38 & -329 & -331 & -348 & 1.6 & 0.0 & 39.3 \\
P11 & 92 & 36 & -326 & -321 & -339 & 1.2 & 0.1 & 37.0 \\
P12 & 86 & 35 & -316 & -313 & -333 & -0.1 & 0.2 & 34.8 \\
P13 & 79 & 36 & -301 & -296 & -317 & -0.3 & 0.1 & 35.3 \\
P14 & 75 & 40 & -285 & -284 & -308 & -2.1 & 0.0 & 37.9 \\
P15 & 74 & 51 & -262 & -265 & -289 & -2.3 & -0.2 & 48.4 \\
P16 & 81 & 64 & -250 & -255 & -283 & -3.5 & -0.5 & 60.7 \\
P17 & 97 & 81 & -247 & -252 & -282 & -4.9 & -0.8 & 76.9 \\
P18 & 124 & 113 & -236 & -251 & -282 & -5.5 & -1.2 & 108.9 \\
\hline
\end{tabular}
\end{table*}

The calculated values of the Peierls energy $\Delta E_P$ in Table \ref{tab:ene}
are fitted to a plane-wave expansion 
which accounts for the periodicity and symmetry as follows:
\begin{eqnarray}
 E(\vec{r}) &=\sum_{\alpha} \left[  C_1 f_e(x_\alpha) + C_2 f_o(x_\alpha)\right] \nonumber \\
&+\sum_{\alpha} \left[ C_3 f_e(2x_\alpha) +C_4 f_o(2x_\alpha) \right]  \nonumber \\
+&C_5 \left[f_e(x_1-x_2)+f_e(x_2-x_3)+f_e(x_3-x_1) \right] ,
\label{eq:pot-fit}
\end{eqnarray}
where $x_\alpha \equiv \vec{r}\cdot \vec{k_\alpha}$, $k_\alpha$ 
points toward three equivalent $<1\bar 10>$directions
with a length $|k_\alpha|= \frac{4\pi}{\sqrt{3} L_0}$ and 
$L_0$ is the distance between adjacent easy-core positions,
 $f_e(x) = (1-\cos (x) )/2$, and  $f_o(x) = \sin (x)/2$.
The coefficients are $C_1=21.82$, $C_2=-14.51$, $C_3=2.59$, $C_4=-2.72$,
and $C_5=-2.89$ meV.
Figure \ref{fig10-pot3d}  shows this fitted potential surface,
with cross-symbols corresponding to
the  values of $\Delta E_P$ in Table \ref{tab:ene}
 their relative values with respect to the fitted values are shown
by perpendicular lines.

\myfig{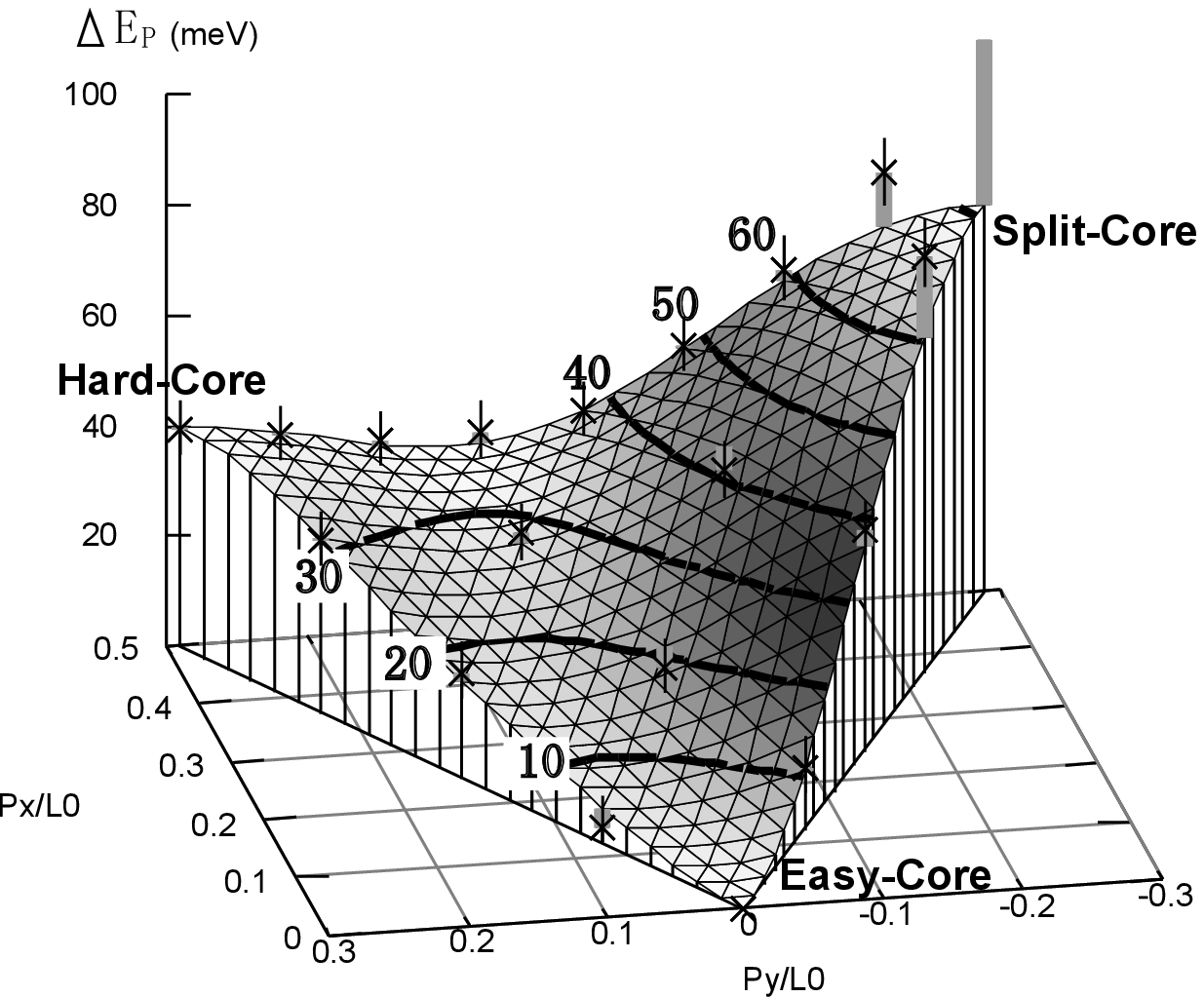}
{Two-dimensional Peierls energy $\Delta E_P$, obtained by the DFT calculations.
The curved surface is a plane-wave expansion fitted to the DFT results.
The cross-symbols are the values of $\Delta E_P$ in Table \ref{tab:ene};
their relative values with respect to the fitted values are shown by gray perpendicular lines.
\fixed{
The thin vertical lines are numerical errors of DFT calculations.}
}

The Peierls energy profile on a line segment between the hard-core and 
the split-core position is shown in Fig. \ref{fig11-ene-saddle},
which is a dividing ridge between two basins of easy-core potential minima.
The minimum in this plot gives a Peierls barrier for dislocation migration,
and one can see that the plot is nearly flat 
between the hard-core position and the M point.
This indicates that the saddle point can be anywhere between these two points.
However, energy minimization of a kink configuration will select the shortest
path between the two easy-core positions and the actual saddle point is expected
to be near the M position.
Fig. \ref{fig11-ene-saddle} also shows the energy profile calculated by
two variants of the embedded atom method (EAM) potential by Mendelev et al. \cite{mendelev03},
which differ from the DFT results significantly.
This difference results in widely different average slip directions,
as will be demonstrated in later sections.
The split-core position,
\fixed{which is metastable in the Mendelev potential,}
is actually unstable as has been shown in Ref. \cite{ventelon-phd}.

Our final estimate for the Peierls barrier 
is $35 \pm 5$meV per $b$, which is in good agreement with
the experimental observation of $27-48$ meV \cite{caillard10,douin09}.
This result is also in good agreement with the DFT 
calculation of SIESTA code, $28-33$ meV \cite{cea-dft-peierls}.


\myfig{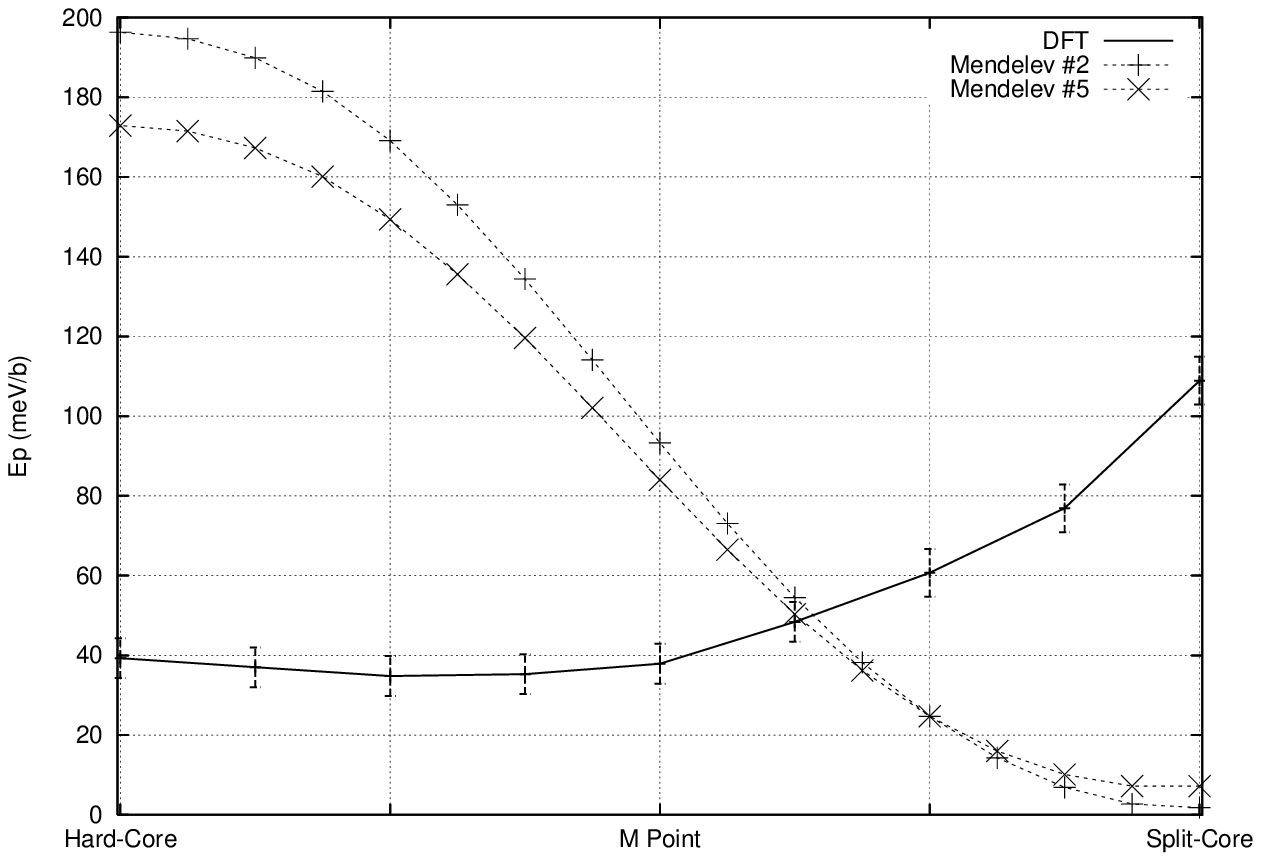}
{Peierls energy profile of a line segment
 between the hard-core and split-core positions calculated by DFT.
Dotted lines show calculation results obtained by
two variants of the EAM potential in Ref.\cite{mendelev03}.
}

Next, we investigate the 
effect of external stress on the Peierls energy.
A uniform shear strain
$\epsilon_{YZ}$ is applied to the system in DFT calculations
by increasing the $Z$ component of the cell vector ${\bf e}_2$ 
by an amount $\epsilon_{YZ} e_2^Y$, where $e_2^Y$ is the Y component of ${\bf e}_2$.
This strain exerts forces in the $X$ direction on the two dislocations.
We investigate four cases,  for the stain  values of
$\epsilon_{YZ}=$ 0\%, 0.5\%, 1.0\% and 1.5\%.
When the two dislocations with opposite helicities move in the $X$ direction by
the same distance, change in the
uniform strain component induced by this movement is canceled out
and the uniform strain remains unchanged. We confirm that the variation in the 
uniform stress $\sigma_{YZ}$ is at most 2\% in the DFT calculation
when the two dislocations move from the easy-core position to the M point.

\fixed{The effect of the applied stress on the 
Peierls energy is given by the Peach-Koehler term as follows:}
\begin{equation}
\{ (\sigma*\vec{b})\times \vec{l}\}\cdot \vec{r}, \label{stress-elas} \label {elaseff}
\end{equation}
where $\sigma*\vec{b}$ is a tensor-vector product of the stress and Burgers vector,
$\vec{l}$ is a vector parallel to the dislocation line and its length is equal to
the dislocation segment under consideration, and $ \vec{r}$ is a dislocation position.
An additional effect of the applied stress, 
on the shape of the Peierls energy $\Delta E_P$,
has been reported in Ref. \cite{rodney09}.
It is important
to check the presence of this additional effect to model the behavior of dislocations
under an applied stress.

When the Burgers vectors of the two dislocations are anti-parallel, 
the term in Eq. (\ref{elaseff}) cancels out and one cannot directly observe it.
Instead, we observe separately the forces $F_x$ and $F_y$ acting on the two dislocations.
By comparing the lattice stress $S_x=-F_x/b^2 \pm \sigma_{YZ}$ and $S_y=-F_y/b^2$
for the different strain cases,
one can see whether the Peierls potential
is affected by the applied stress or not.

Fig. \ref{fig12-stress-eta} shows a schematic of the method of calculation for
the lattice stress.
Since the applied stress breaks the symmetry of inversion in the $X$ direction,
the lattice stress curve plotted against the 
reaction coordinate $\eta$ can be asymmetric around $\eta=1/2$,
as shown by the solid line in Fig. \ref{fig12-stress-eta}.
For the  ``plus'' dislocation,
its position between the easy-core and the adjacent M point
which is on the $+X$ direction corresponds to the range  $0\leq \eta \leq 1/2$,
while for the ``minus'' dislocation it corresponds to $1/2 \leq \eta \leq 1$.
Observing the forces acting on the two dislocations separately, we can
obtain the complete lattice stress curve.

\myfig{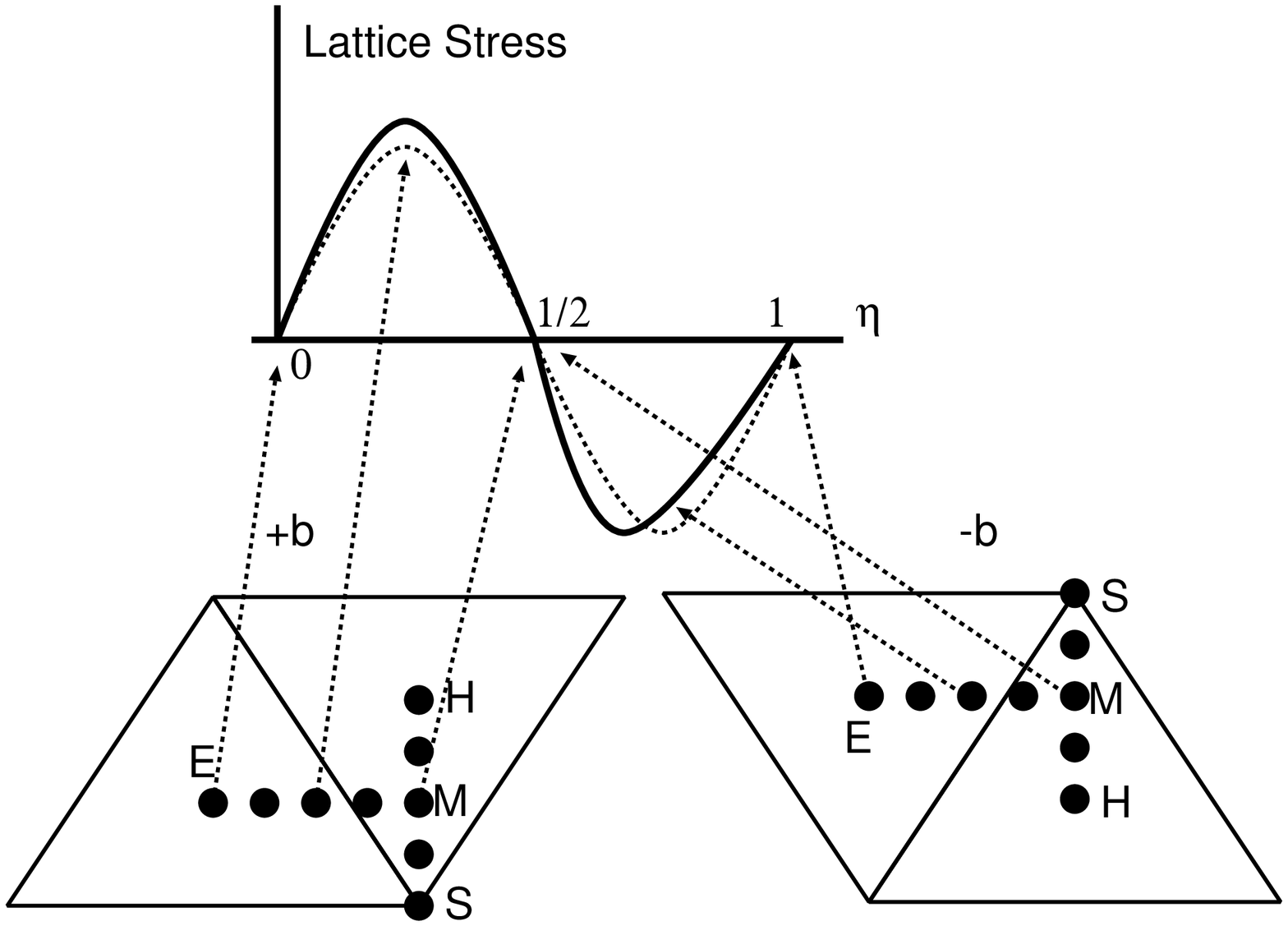}
{
Schematic of calculation of lattice stress curve 
    from the forces acting on the two dislocations with opposite helicity.
    Black circles denote the dislocation positions where the lattice stress
    is observed in the DFT calculations.
Solid and dotted curves represent the lattice stress of the zero stress and
non-zero stress cases, respectively.
Both dislocations are moved in the same direction, and  the forces
on dislocations with the positive and negative helicities are used to calculate
the lattice stress for $0\leq \eta \leq 1/2$ and $1/2 \leq \eta \leq 1$,
respectively.
}

Figure \ref{fig13-sxsy}
shows a comparison of the lattice stress for different applied shear strains.
Observed values of the uniform stress $\sigma_{YZ}$ 
which are subtracted from the force
for each strain case are also shown by horizontal lines.
Although the curves do not conform to a single line,
the deviation of $S_x$ is best described by shifting to the $-\eta$ direction
as the strain is increased,
while the shape of the curves of $S_y$  changes slightly for strains greater than 1.0\%.
If the applied stress only  shift the curve and does not change its shape,
the shape of $\Delta E_P$ remains unchanged.
Fig. \ref{fig14-ene-sd2}
shows a comparison of the Peierls energy profile of a line segment between the 
hard-core and split-core
positions for the $\epsilon_{YZ}= 0$ and 1.5 \% cases.
The difference between the two cases is 
less than $5$ meV in the saddle point region.

\myfig{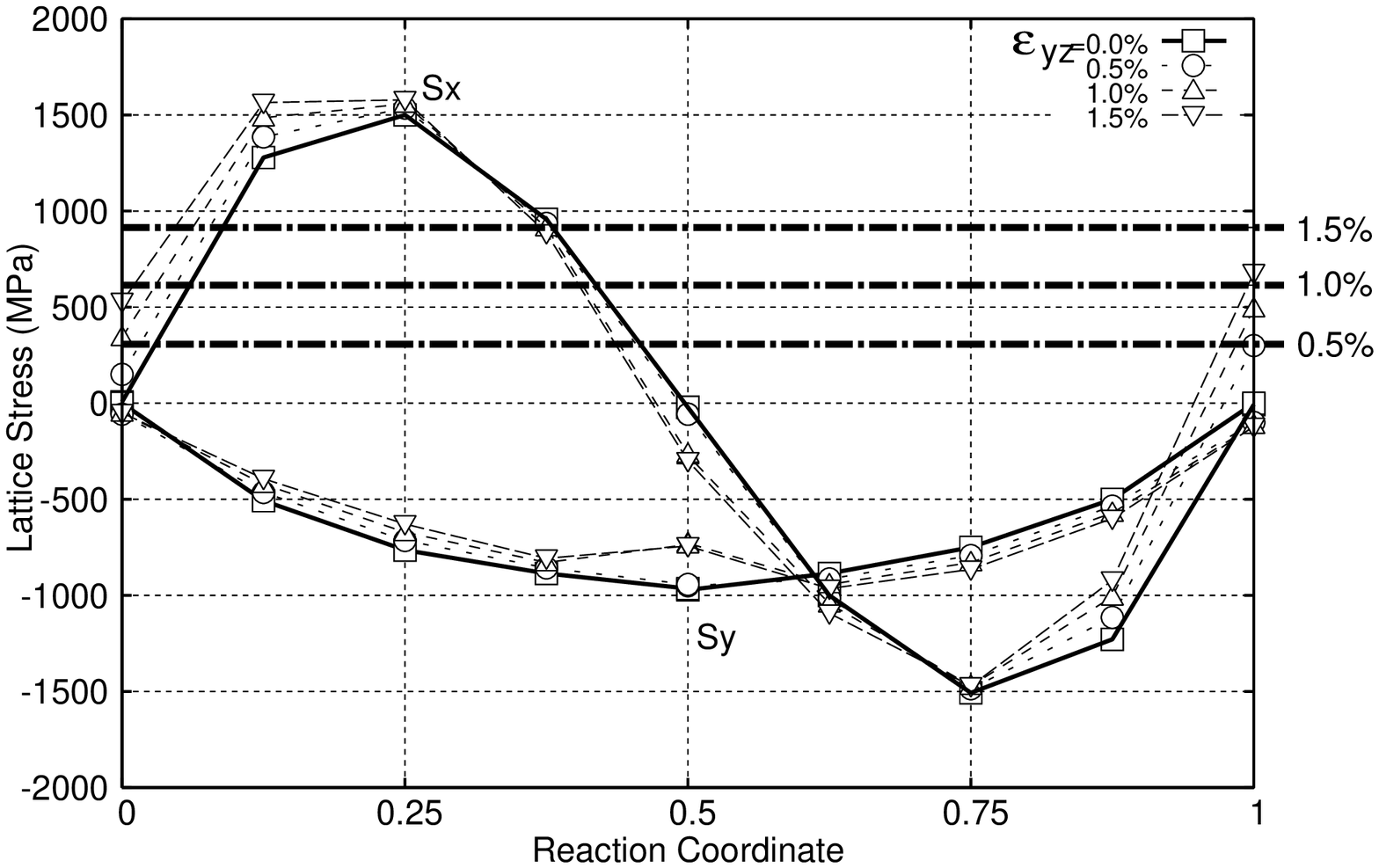}
{
Comparison of the lattice stress for different applied shear strains of 
 $\epsilon_{YZ}=$ 0\%, 0.5\%, 1.0\% and 1.5\%.
The uniform stress $\sigma_{YZ}$ observed in the DFT simulation
for each strain is shown by horizontal dashed lines.
}
\myfig{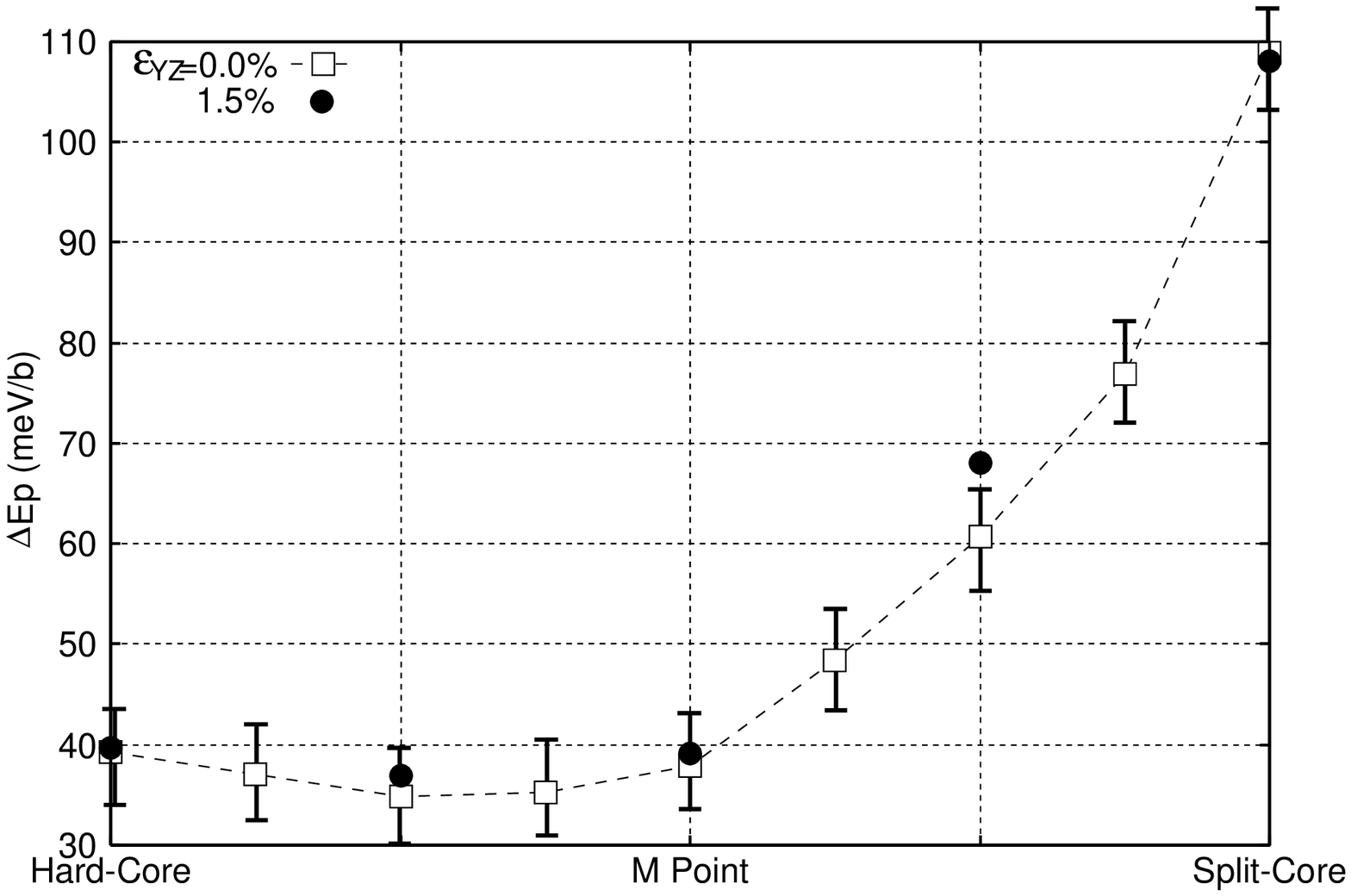}
{
Comparison of  Peierls energy profile on a line segment between the hard-core and the split-core
    position for  $\epsilon_{YZ}= 0$ and 1.5 \% cases.
}

From these results,
we conclude that the effect of the applied stress is described well
by the  linear elasticity  theory given by Eq. \ref{elaseff}
and its additional effect on the shape of 
$E_P$ is very weak, at least up to
 $900$ MPa, which is the maximum stress investigated in the present work.
The Peierls stress, 
above which the Peierls energy landscape has no stable minima,
is estimated from Eq. (\ref{eq:pot-fit}) to be $1000\pm 50$ MPa.
This stress is far stronger than the experimental estimation of $\sim 400$ MPa
(such a discrepancy between atomistic models and experiment
is a general tendency in bcc metals \cite{pileup,domain05,Ta-peierls-stress,groger08}).
The Peierls stress is determined by the maximum
slope of the potential energy, and one can estimate its lower bound
by  $\Delta E/\Delta x b^{-2}$, where $\Delta E$ is the saddle point energy
and $\Delta x$ is the distance between
the easy-core position and the saddle point. 
Since the value of $\Delta E$ agrees well with the experimental estimation
and the lower bound Peierls stress calculated with $\Delta E=30$ meV gives $690$ MPa, which
is still far stronger than the experimental estimation,
this discrepancy of the Peierls stress cannot be attributed to
the shape of the potential energy landscape but
should be attributed to other phenomena,
such as interactions between dislocations \cite{pileup}.

\section{Line Tension Model}
The dislocation mobility is determined by the formation
\fixed{enthalpy} of a dislocation kink,
a defect at which a dislocation line
moves from one Peierls energy minimum to the next \cite{domain05}.
Fig. \ref{fig15-totan} depicts 
the kink nucleation mechanism of dislocation migration. 
As shown in the inset of the figure, the total
dislocation \fixed{enthalpy} reaches a maximum when some part of the dislocation
line bulges and overcomes the Peierls barrier. After that, a fully formed
kink pair moves in the opposite direction, lowering the total \fixed{enthalpy}
as they separate until the whole
dislocation line moves to the next minimum of the potential energy.

\myfig{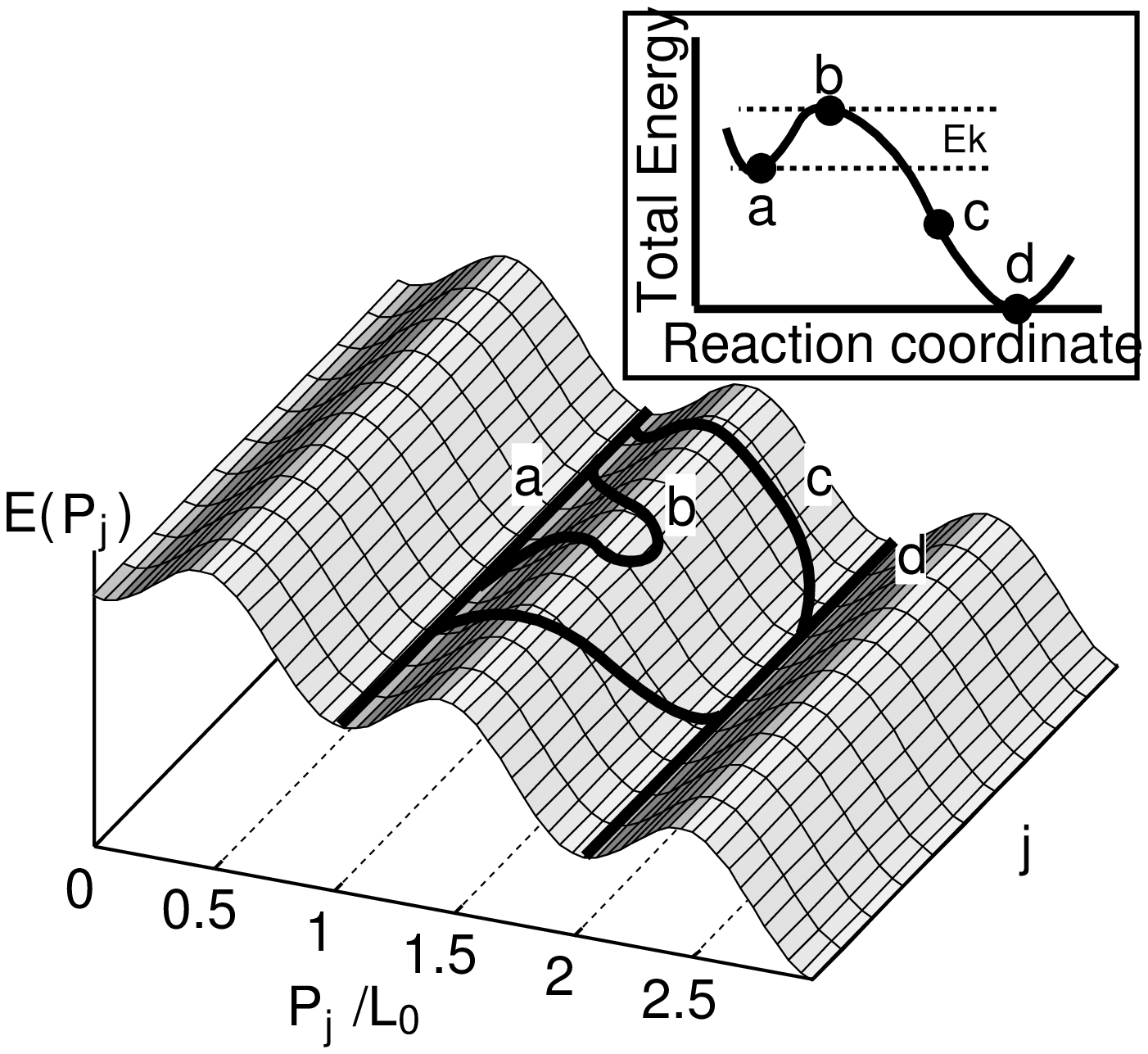}
{
Schematic  of dislocation movement by
the kink nucleation  mechanism.
(a) A straight dislocation lies in a valley of Peierls potential energy.
(b) Some part of the dislocation bulges and overcomes
the Peierls barrier.
(c) A kink pair is fully formed and two kinks move in the opposite direction, lowering
    the total \fixed{enthalpy} as they separate.
(d) Both kinks are absorbed at the boundary of the dislocation,
completing the movement process.
}

\fixed{The kink formation enthalpy
calculated from the molecular statics \cite{kink-rodneymd}
or line-tension model adjusted on atomistic
calculations \cite{rodney09,lt-edagawa97}
has been shown to reproduce the dislocation velocity
in the molecular dynamics simulations.
For a screw dislocation in bcc metals, the kink width
is estimated as $10-20 b$ \cite{md-kink},
and a direct calculation of kink formation 
enthalpy requires at least several tens of slabs
which amount to thousands of atoms.
Since plane-wave basis DFT of thousands of atoms
is not feasible,  we use the following line tension model
to estimate the kink formation enthalpy:} 

\begin{eqnarray}
E_{LT}=&{K \over 2}\sum_j (\vec{P}_j - \vec{P}_{j+1})^2 \nonumber \\
       &+ \sum_j \Delta E_P(\vec{P}_j) +
\{ (\sigma*b)\times \vec{l}\}\cdot \vec{P}_j  \label{equ:ltene} ,
\end{eqnarray}
where
K is a constant related to the elastic constants of the material,
$j$ is an index of thin slabs with the thickness $b$  parallel to the $Z$ axis,
$\vec{P}_j=(P_j^x, P_j^y)$ is the  position of a dislocation core in the slab $j$, 
$\Delta E_P$ is the Peierls barrier per Burgers vector $b$ obtained in the
previous section, and the third term is the contribution from the external stress.

The constant K can be calculated by
comparing the quadratic expansion of Eq. \ref{equ:ltene}
in $P_j^x$ and $P_j^y$ with the expansion
of $E_d(x^d +\delta)$ in $\delta$, which is given as follows:
\beq
E_d(x^d+\delta)-E_d(x^d)\sim
-\sum_i \delta_i F_i + \frac{1}{2}\sum_{ij} H_{ij} \delta_i \delta_j.
\eeq
The perturbation in $x^d$ also changes $\bar{x}^o(x^d)$, but
the change in $x^o$ does not affect the energy.
The matrix $H_{ij}$ can be calculated from the change in the forces acting on each $x^d$
as follows:
\beq
F_i(x^d+\delta)-F_i(x^d)= \sum_j H_{ij} \delta_j. \label{GijExp}
\eeq

First we demonstrate the calculation scheme of $K$
using an EAM potential, and  compare the kink shape predicted by
the line tension model with the actual kink shape obtained in MD simulation.
An isolated screw dislocation
of length $60b$ is provided
using the system consisting of 1452 atoms per layer.
With a Mendelev potential of the variant 5 \cite{mendelev03},
this system is
fully relaxed  to the easy-core configuration while the outermost atoms 
are fixed to the linear elasticity solutions.
Then one atom in the innermost three atom columns is displaced in the $+Z$ direction
by $0.01$ \AA, and a force acting on every atom is calculated using the EAM potential.
The induced force
on the two atoms which are located  directly above and below the displaced atom,
is found to be $29$ meV/$\AA$ in the $Z$ direction,
while that on the displaced atom itself is $-82$ meV/$\AA$ in the same direction.
Forces induced on the other atoms are negligible. Thus $H_{ij}$ is calculated
to be $H_{ii}=8.2$ eV/$\AA^2$
for the diagonal element and $H_{ij}=-2.9$ eV/$\AA^2$
for the neighboring pair on the same atomic row.
If we denote the $Z$ displacement of the innermost three atoms on the layer $j$
by $z_{j,1}$, $z_{j,2}$ and $z_{j,3}$, the total energy can be expanded in
these quantities as
\beq
\frac{1}{2} \sum_j \sum_{a=1,2,3} A z_{j,a}^2 +K_0 (z_{j,a}-z_{j-1,a})^2,
\eeq
with $A=2.4$ eV/\AA and $K_0=2.9$ eV/\AA.
Together with Eqs. (\ref{pxpy}) and (\ref{equ:ltene}),
 we get $K=\frac{9}{32} K_0=0.816$ eV/\AA$^2$
for the Mendelev potential.
The Peierls energy landscape of the same potential is calculated using a single-layer,
isolated dislocation configuration, and fitted to the form of Eq. \ref{eq:pot-fit}.
The coefficients are 
$C_1=40.4$, $C_2= 66.9$, $C_3= -0.1$, $C_4= 2.3$ and $C_5= 0.6$ meV.

Fig. \ref{fig16-kinkmd} shows a comparison of kink shapes of the
MD simulation and the line tension model with these parameters.
\fixed{In addition to the
good agreement in the kink shape in the $x$ direction,
which was  shown in previous studies using the
line tension model \cite{rodney09},
the kink shape in the $y$ direction also shows excellent agreement.}
The kink pair energy calculated by the MD relaxation 
is $0.78$ eV, while  the line tension model
gives  $0.69$ eV. It has been shown that
the kink shape and energy is asymmetric between the two types of kinks
in the MD calculations \cite{md-kink}. This means that there should be
higher-order terms of the gradient $\vec{P}_j-\vec{P}_{j-1}$ in the energy,
and the difference of the energy between the MD and the line tension model
can be attributed to this term.
As a whole, these results clearly demonstrate the validity of the calculation
scheme of $K$ presented above.
By a simple scaling argument, one can show that the kink formation energy
is proportional to $\sqrt{K E_P}$, and fortunately the uncertainty in $K$
only modestly affect the estimate of kink formation energy.
\myfig{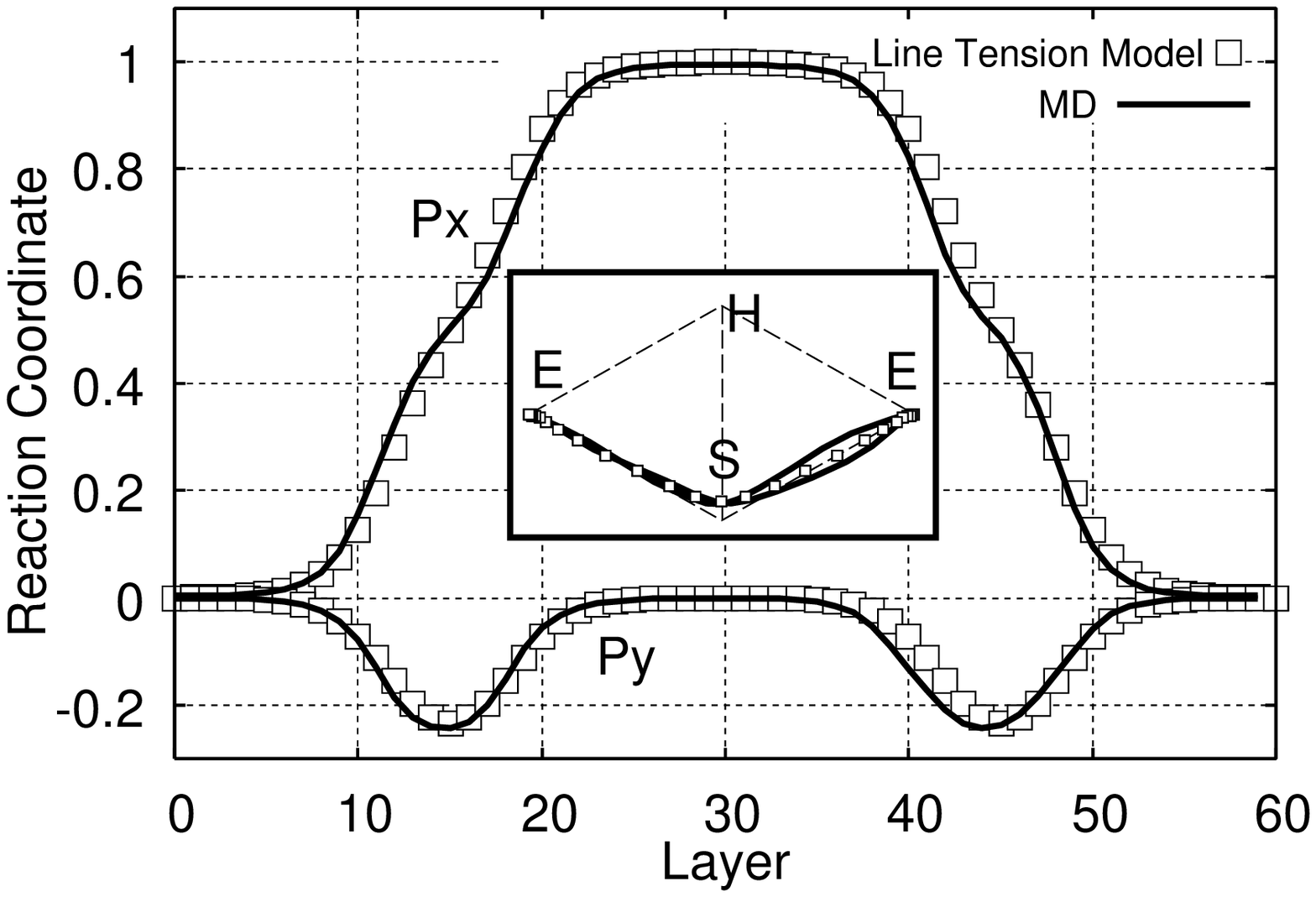}
{
Comparison of the kink shapes of the MD simulation
and the line tension model. $P_x$ and $P_y$ are normalized 
by the distance between two easy-core positions.
The inset shows the two-dimensional kink shape in the XY plane
for both cases.
}

To estimate the value $K$ in the DFT calculation, a single-layer hexagonal 
system consisting of 108 atoms, which contains a single easy-core configuration 
in the center, is prepared and relaxed, while the outermost 33 atoms are fixed.
The obtained configuration planes are then stacked on each other
to form a three-layer system,
and one of the innermost atoms is displaced in the $Z$ direction
by $0.01 \AA$.
The forces induced by this displacement are calculated using the DFT.
From the MD result and the DFT calculation of the perfect crystal,
we found that forces on the second or farther neighboring atoms can be ignored,
and that a three-layer system is suffice to calculate the matrix $H_{ij}$.
The force induced on the two atoms
located 
directly above and below the displaced atom is found to be $0.0308$ eV/$\AA$;
the forces on the atoms in the other atomic column are negligible.
From this result,
we get $K_0=3.08$ eV/\AA$^2$, and consequently $K=0.866$ eV/\AA$^2$.

The kink pair nucleation enthalpy $E_k$ and its dependence 
on the applied stress is calculated by
applying the transition state search method \cite{stringmethod} 
to the model (\ref{equ:ltene}).
Fig. \ref{fig17-kinkdft} shows
the kink shape predicted from the line tension model,
when there is no applied stress.
The kink pair energy and the kink width are estimated to be $0.73$ eV
and $10b$, respectively, which are in good agreement with previous studies \cite{md-kink}
and an experimental estimate \cite{spitzig70b}.
\fixed{Since the kink width is proportional to $b \sqrt{K/E_P}$
and $E_P$ is much higher than that of the Mendelev potential, the kink
    width is much shorter compared to the Mendelev potential case.}

\myfig{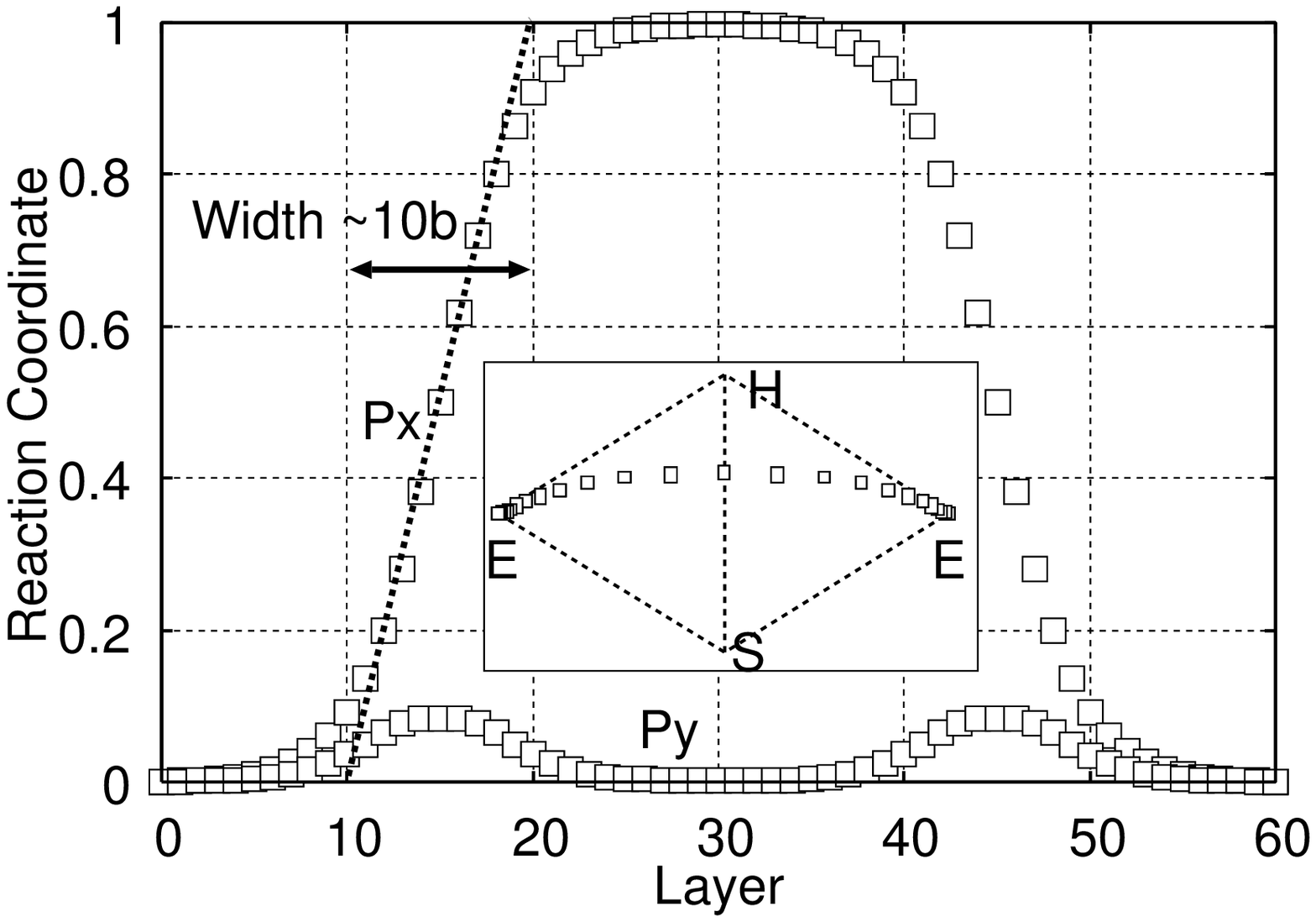}
{
Kink shape obtained from the line tension model with
the parameters given by the DFT calculations.
$P_x$ and $P_y$ are normalized 
by the distance between two easy-core positions.
The kink width is estimated to be \fixed{$10b$}.
The inset shows the two dimensional kink shape in the XY plane.
}

Fig. \ref{fig18-ekink} shows the dependence of $E_k$
on the applied stress for three maximum resolved shear stress directions,
$[111](1\bar{1}0)$, $[111](11\bar{2})$ (twinning), and  $[111](2\bar{1}\bar{1})$ (anti-twinning),
together with  experimentally observed values from  Ref.\cite{spitzig70b} and their linear fit.
For the case of stress in the
$(1\bar{1}0)$ plane, the direction of a dislocation migration coincides
with the stress direction,
while for the case of twinning and anti-twinning, the high Peierls barrier around the 
split-core position prohibits a
direct migration in the $\{211\}$ planes and the macroscopic slip in the $\{211\}$ planes is
realized by random successive migrations in the two equivalent $\{110\}$ planes,
assuming that the screw dislocation is terminated by either a free surface or a
periodic boundary.
A trace of this random migration has been experimentally observed as a wavy pattern on a
free surface of a thin film \cite{caillard10}.
When the screw dislocation is part of a loop, the edge part energetically favors
a straight line and random migration is expected to be suppressed.
The overall tendencies shown in Fig. \ref{fig18-ekink}
such as  the non-Schmit behavior 
that a slip in the twinning direction is easier than in the anti-twinning direction,
agrees with the experiment in Ref.\cite{spitzig70b}, except for the already discussed discrepancy
of the Peierls stress.

\myfig{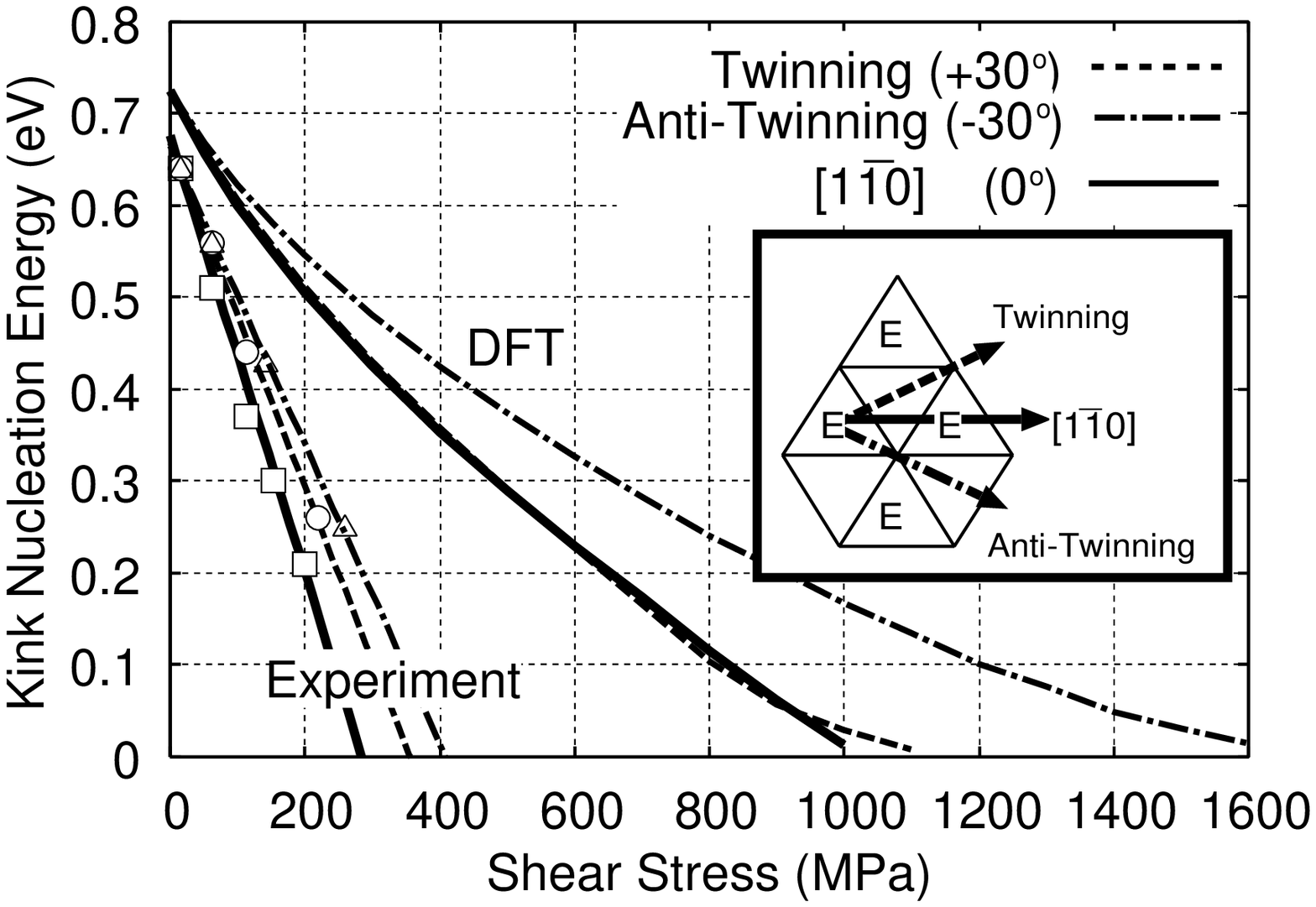}
{
Stress dependence of the kink nucleation enthalpy
for three stress directions,
 $[111](1\bar{1}0)$,
 $[111](11\bar{2})$ (twinning),
 and $[111](2\bar{1}\bar{1})$ (anti-twinning).
Experimental data from Ref. \cite{spitzig70b} are also shown as
symbols, together with  straight line fits.
}

The average slip direction at a finite temperature
is calculated using the
Brownian dynamics simulation:
\begin{eqnarray}
\vec{P}_j (t+\Delta t) = \vec{P}_j - \Delta t \nabla_j  E_{LT} 
+ \sqrt{2 k_B T \Delta t} \vec{\eta}_j
,
\label{equ:lang}
\end{eqnarray}
where $\nabla_j$ denotes a derivative with respect to $\vec{P}_j$,
$T$ is the temperature, $ k_B$ is the Boltzmann constant
and $\vec{\eta}_j$ is a two-component
Gaussian random variable whose average and variance are 0 and 1, respectively.
It should be noted that the time scale in Eq. (\ref{equ:lang}) does not correspond to
the physical one, 
and any momentum effect 
is not incorporated.
However, Eq. (\ref{equ:lang}) is able to capture 
the entropic effect on the migration frequency and 
the interaction between  kinks.
A time step of $\Delta t = 1.1 \times 10^{-3}$ is used,
which ensures that the change in $\vec{P}_j$ at each step is
at most 3\% of the distance between two easy-core positions,
and the iteration is repeated $10^5$ times for each case.


Fig. \ref{fig19-slip-dir} shows the dependence of the average slip direction  
on the stress and temperature, for the case of shear stress 
in the $[111](1\bar{1}0)$ direction.
The angle is relative to the $(1\bar{1}0)$
plane and is defined as 
positive for the twinning direction.
\fixed{ Similar plots have been obtained based on the
MD simulations \cite{md-slip-dir,marian04}, and the main difference between
the present result and the MD simulations is the magnitude of 
the deviation from the $(1\bar{1}0)$ plane.
For all the cases studied here, it is less than 2 degrees,}
in contrast to the MD results, in which the deviation is as large as 20 degrees.
Experimental observation of dislocation motion
\cite{caillard10} indicates that 
the slip plane for the case of maximum resolved shear stress
in the $<111>(1\bar{1}0)$ direction
is primarily $(1\bar{1}0)$,
and we conclude that the deviation of the slip direction in MD is
an artifact of the EAM empirical potential.
The three lines  $C1$, $C2$ and $C3$ in Fig. \ref{fig19-slip-dir},
which separate the  regions of different slip behavior,
are given by equations
$T= E_k ^0/20 k_B$, $T=(E_k^+-E_k^0)/2k_B$ and $T=(E_k^--E_k^0)/2k_B$, respectively,
where $E_k ^0$, $E_k^+$, and $E_k^-$ denote the kink nucleation enthalpy
to the neighbor easy-core configuration located in the
directions of $0$, $+60$ and $-60 ^\circ$, respectively, for
a given value of applied shear stress in the $[111](1\bar{1}0)$ direction
calculated using the line tension model.
Below $C1$, the kink nucleation is so rare that we cannot obtain sufficient data.
Above $C2$, nucleation in the $+60^\circ$  direction becomes
comparable to that in the $0^\circ$  direction and the average slip
direction becomes positive. When the applied stress is greater than 800MPa,
the kink nucleation in this direction becomes unstable and eventually
becomes a kink pair in the $0^\circ$ degree direction, and the curve $C2$
becomes a vertical line at  800MPa.
Above $C3$, kink nucleation in the $0$, $+60$ and $-60 ^\circ$
direction all becomes comparable and the average slip direction decreases as the
temperature is increased.

\myfig{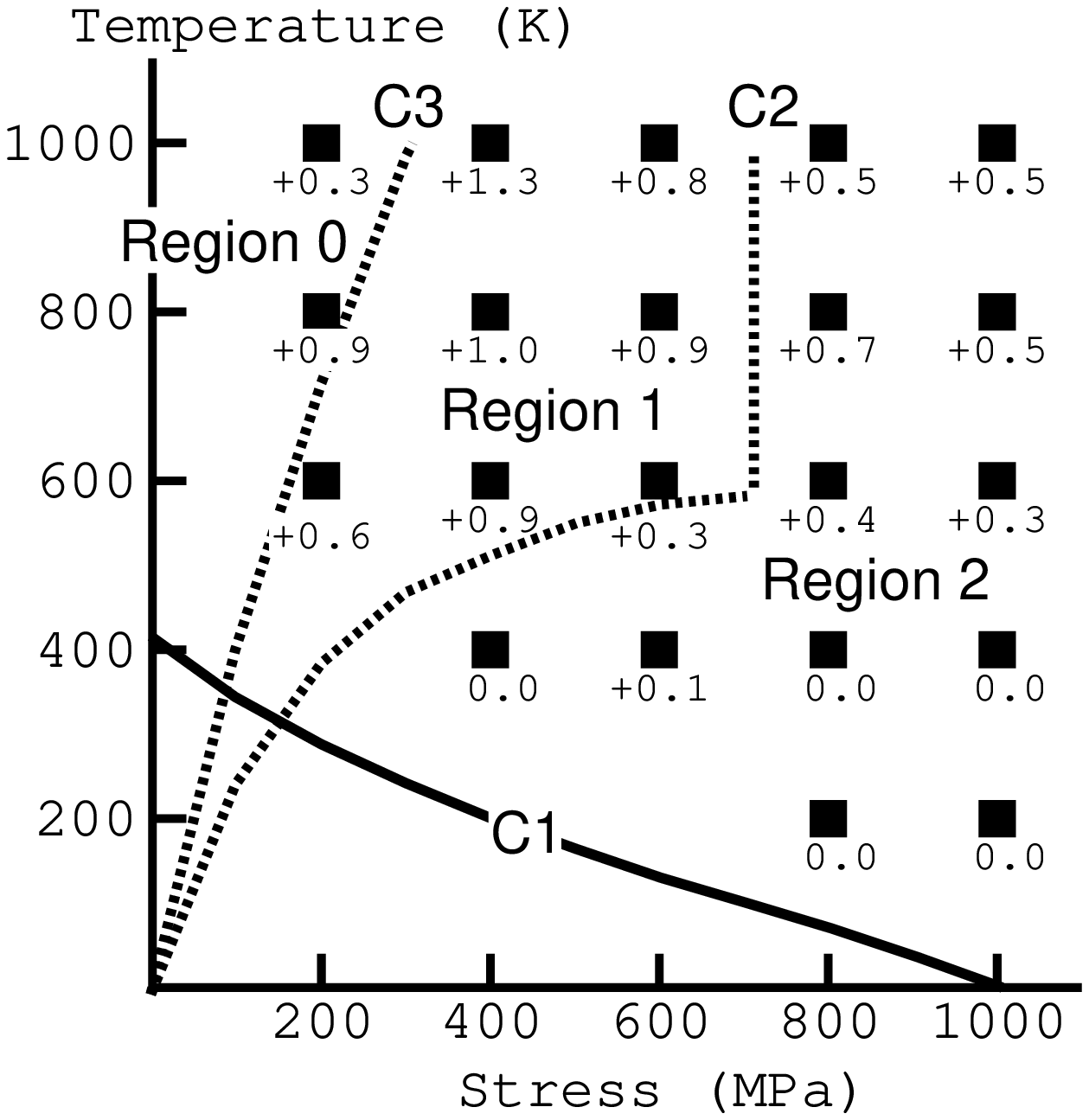}
{
Temperature and stress dependence of
the average slip direction, relative to 
the maximum resolved shear stress direction of $[111](01\bar{1})$.
The curves $C1$, $C2$ and $C3$ are the boundaries between
different slip behavior regions, calculated from the difference
between kink nucleation energies in different directions.
See the main text for  details.
}

Fig. \ref{fig20-mig-schem}
shows how the applied stress affect the nucleation energies $E_k ^0$, $E_k^+$, and $E_k^-$.
In the present study, migration paths
are close to the straight line and the contribution of the applied stress $\sigma$
to the Peierls barrier is approximately  proportional to $\sigma \cos(\theta) \chi$,
where $0\leq \chi \leq 1$ denotes a reaction coordinate of the dislocation migration
and $\theta=$ $0$, $+60$ and $-60^\circ$.
The migration path for the $+60 ^\circ$ case can bend toward 
the energetically favorable direction
and the migration energy is further lowered,
while for the $-60 ^\circ$ case the high energy of the split-core
position prevents the path from bending to the energetically favorable direction
and the migration energy is slightly higher than the $+60 ^\circ$ case.
Thus we have $E_k ^0 < E_k^+ < E_k^-$, in which case
the stress and temperature dependence of the slip direction in 
Fig. \ref{fig19-slip-dir} is 
reproduced.

For the case of empirical EAM potential,
the migration path passes though the split-core positions
and the effect of the applied stress differs significantly from the
present study case, as shown in Fig. \ref{fig20-mig-schem} (b).
The migration energes for the $0^\circ$ and  $-60^\circ$ cases
are almost the same, since the saddle point of the migration
for these two cases are  close together even though the corresponding
atomic configurations differ.
Consequently, the average slip direction becomes close to $-30^\circ$
as soon as the temperature is high enough to overcome the 
small energy difference between the $0^\circ$ and   $-60^\circ$
migrations.
The double-humped potential shape also
causes discontinuity in the kink nucleation \fixed{enthalpy} at a certain
applied stress \cite{gordon10}.


\myfig{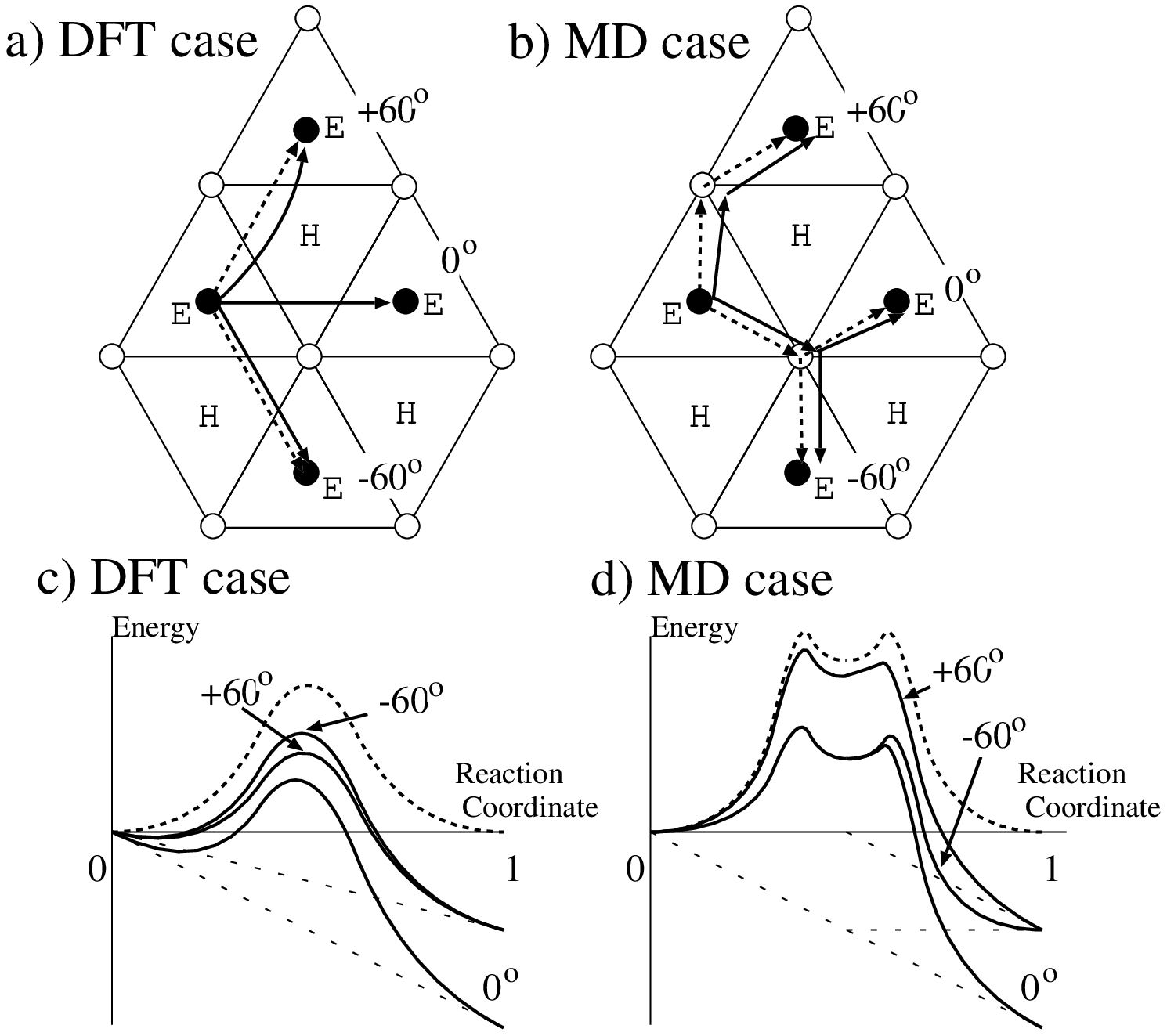}
{
Schematic of the effect of the applied stress
 on the average slip direction for the cases of
 (a) the present study and (b) the empirical EAM potential.
Solid and dotted lines show migration pathways with and without
the applied stress, respectively.
(c and d)  Schematic behavior of the Peierls
barrier for both cases. See the main text for details.
}

\section{Summary and Conclusion}
Essential properties of a 
screw dislocation in bcc iron have been calculated and determined
 by the plane-wave-based DFT calculations.
The properties of dislocation kinks are determined using a line tension model 
which is based on the DFT results.
Our findings are summarized as follows:
 (i) a screw dislocation migrates from one easy core position to an 
 adjacent easy core position, following a nearly straight path;
 (ii) the saddle point of the migration is around a midpoint of 
 two easy core positions, at which the Peierls barrier
is estimated to be $30-40$ meV;
 (iii) a kink pair nucleation enthalpy is estimated to be
 $0.73$ eV when no external stress is applied;
 (iv) when the maximum resolved shear stress is in 
 the $<111>\{1\bar{1}0\}$ direction,
 the average slip direction remains in the $\{1\bar{1}0\}$
 plane with a deviation of less than 2 degrees;
 (v) when the maximum resolved shear stress is in the $<111>\{\bar{2}11\}$ direction,
   two equivalent $\{1\bar{1}0\}$ slips are activated randomly;
 and (vi) the Peierls stress is estimated to be $1000$ MPa,
 with a lower bound of $620$ MPa.

These results are consistent with experimental estimation, except for the far larger
value  of the Peierls stress (vi)
compared to the experimental estimate. This discrepancy in the Peierls
stress between atomistic scale calculations and experiments is a general tendency in bcc metals,
and we foresee that
some mesoscopic modeling is required, such as the pile-up model proposed by 
Gr\"{o}ger et al.\cite{pileup},
to relate the calculated Peierls stress to the
experimentally estimated yield stress in the zero temperature limit.
Points  (iv) and (v) are  in clear contrast to the molecular dynamics simulations of 
bcc iron \cite{md-slip-dir}.
This is because the fundamental properties of migration for a screw dislocation
 have not been included  in the fitting targets in the previous development
 of empirical potential in bcc metals.
Recently, an improved EAM potential has been developed by Gordon et al. \cite{gordon11}
in an effort to reproduce both the symmetric easy-core structure and
the single-humped Peierls energy. Gordon et al.
 concluded that conventional EAM potential is not capable of reproducing 
 these two properties simultaneously.
Further improvements in EAM potential
based on  the dislocation migration properties obtained in the present work,
possibly by including the anisotropic terms or the
magnetic degrees of freedom \cite{pot-mag}  in the potential, 
are strongly expected for the reliable simulation of plastic deformation in bcc iron.

\section*{Acknowledgments}
We thank  F.~Willaime, L.~Proville, L.~Ventelon, E.~ Clouet, and J.~Marian
for helpful discussions.

\section*{Appendix A: Anisotropic elasticity}
Owing to the elastic anisotropy,
the elastic constants in the Cartesian coordinate in Fig. \ref{fig03-cell}
change from the cubic axis case by amounts 
$\Delta C_{xx,xx}=\Delta C_{yy,yy}=-A/2$,
$\Delta C_{zz,zz}=-2A/3$,
$\Delta C_{yy,zz}=\Delta C_{zz,xx}=A/3$,
$\Delta C_{xx,yy}=A/6$,
$\Delta C_{zx,zx}=\Delta C_{yz,yz}=A/3$,
$\Delta C_{xy,xy}=A/6$,
$\Delta C_{xx,zx}=\Delta C_{xy,yz}=-\frac{A}{3\sqrt{2}}$,
and $\Delta C_{yy,zx}=\frac{A}{3\sqrt{2}}$,
where $A=C_{11}-C_{12}-2C_{44}$ is the anisotropy parameter.
The isotropic linear elasticity solution 
for the $<111>$ screw dislocation has component
$\epsilon_{yz}$ and $\epsilon_{xz}$, 
and induces excess stress field
$\sigma_{xx}=-A^\prime \epsilon_{zx}$,
$\sigma_{yy}=A^\prime \epsilon_{zx}$, and
$\sigma_{xy}=-A^\prime \epsilon_{yz}$ where $A^\prime=\frac{A}{3\sqrt{2}}$.
However, these excess stress fields satisfy the equilibrium condition
$\sum_b \partial_b \sigma_{ab}=0$ and the solution itself is identical
to the isotropic case.

\end{document}